\documentclass[%
preprint,
showpacs,
amsmath,amssymb,
aps,
prb,
]{revtex4-1}
\usepackage{graphicx}
\usepackage{dcolumn}
\usepackage{bm}
\usepackage{color}
\usepackage{verbatim}

\begin{document}

\preprint{APS/123-QED}

\title{Effects of low-lying excitations on ground-state energy and energy gap of Sherrington-Kirkpatrick model in transverse field}

\author{Yang Wei Koh}
\email{patrickkyw@gmail.com}
\affiliation{Bioinformatics Institute, 30 Biopolis Street, No. 07-01, Matrix, Singapore 138671}

\date{\today}

\begin{abstract}
We present an extensive numerical study of the Sherrington-Kirkpatrick model in transverse field. Recent numerical studies of quantum spin-glasses have focused on exact diagonalization of the full Hamiltonian for small systems ($\approx$ 20 spins). However, such exact numerical treatments are difficult to apply on larger systems. We propose making an approximation by using only a subspace of the full Hilbert space spanned by low-lying excitations consisting of one-spin flipped and two-spin flipped states. The approximation procedure is carried out within the theoretical framework of Hartree-Fock approximation and Configuration Interaction. Although not exact, our approach allows us to study larger system sizes comparable to that achievable by state of the art Quantum Monte Carlo simulations. We calculate two quantities of interest due to recent advances in quantum annealing, the ground-state energy and the energy gap between the ground and first excited state. For the energy gap, we derive a novel formula that enables it to be calculated using just the ground-state wavefunction, thereby circumventing the need to diagonalize the Hamiltonian. We calculate the scalings of the energy gap and the leading correction to the extensive part of the ground-state energy with system size, which are difficult to obtain with current methods.

\end{abstract}
\maketitle


\section{Introduction}
\label{sec.introduction}

The study of quantum spin-glass has a long history dating back to the early seminal work of Bray and Moore on the random quantum Heisenberg model \cite{Bray80}. Many different models \cite{Yamamoto87, Thirumalai89, Buttner90, Goldschmidt90b, Usadel91, Goldschmidt90, Dobrosavljevic90, Obuchi07, Ye93, Read95, Laumann08, Krzakala08} as well as theoretical methods \cite{Bray80,Yamamoto87, Thirumalai89, Buttner90, Goldschmidt90b, Usadel91, Goldschmidt90, Dobrosavljevic90, Obuchi07, Ye93, Read95, Laumann08, Krzakala08,Takahashi07} for studying them have since been proposed. In most of these earlier works, the approach is mainly a mean field one based on a combination of Suzuki-Trotter decomposition and replica theory \cite{Bray80}. More recently, quantum spin-glasses have attracted attention within the context of quantum annealing \cite{Kadowaki98,Das08} and adiabatic quantum computation \cite{Farhi01}. Here, the quantity of interest is the energy gap between the ground and first excited state as it determines the success rate of the annealing process \cite{Sei07, Morita08}, especially in the thermodynamic limit \cite{Jorg10, Ohzeki11}. To compute the energy gap, precise calculation of the energies of the lowest two energy levels is necessary. Earlier mean field approaches are no longer sufficient since they give only the ground-state energy, and even that is strictly speaking only correct in the thermodynamic limit. Furthermore, the first excited state arises from the excitation of just a few spins and is very close in energy to the ground-state. The energy gap is therefore very small compared to the ground-state energy and is not an extensive quantity. One usually resorts to numerical methods when computing the energy gap in quantum spin-glasses \cite{Jorg08, Young08, Young10, Takahashi10a, Takahashi10b}.

In practice, one is frequently interested in the behavior of the energy gap when the number of spins is large, and in this respect the most important difficulty faced in the numerical study of quantum spin-glasses is the apparent lack of conserved quantities, i.e. operators that commute with the Hamiltonian. This is in stark contrast to, for instance, a non-disordered spin system \cite{Jorg10, Seki12, Seoane12} such as the infinite range ferromagnetic Ising model in transverse field
\begin{equation}
H_{\mathrm{ferro}}=-\frac{J^{\prime}}{N}\left(\sum_{i=1}^{N}\sigma_i^{z}\right)^2 -\Gamma^{\prime} \sum_{i=1}^{N}\sigma_i^x,
\label{eq.I.01}
\end{equation}
where $\sigma_i^{\alpha}$ $(\alpha=x,y,z)$ is the $\alpha$-direction Pauli matrix of the $i$th spin, $N$ is the total number of spins, and $J^{\prime}$ and $\Gamma^{\prime}$ are, respectively, the strengths of the ferromagnetic coupling and transverse field. For $H_{\mathrm{ferro}}$, the total angular momentum 
\begin{equation}
S^2=
\left(\sum_{i=1}^{N}\sigma_i^{x}\right)^2 
+
\left(\sum_{i=1}^{N}\sigma_i^{y}\right)^2
+
\left(\sum_{i=1}^{N}\sigma_i^{z}\right)^2
\label{eq.I.02}
\end{equation}
is a conserved quantity and the Hamiltonian matrix takes a block diagonal form. In particular, the ground and first excited states lie in the block with the largest total angular momentum, and it is possible to diagonalize this block matrix for large $N$ because its dimension scales only linearly with $N$. Fig. \ref{fig.Fig00} shows the energy gap of $H_{\mathrm{ferro}}$ computed in this way for $N=20, 250$, and 2000. Unfortunately, $S^2$ is no longer conserved for quantum spin-glasses, and one must work with the full Hilbert space whose dimension scales exponentially with $N$.

There are currently three main approaches for computing the low-lying energy levels of quantum spin-glasses. The first is to diagonalize the full Hamiltonian matrix using standard numerical routines such as Jacobi's method or Householder reduction \cite{Press07}. All energy levels are obtained, but the size of the matrices and computational time involved mean that this approach is practical only for relatively small systems ($N\approx10$). The second method is Lanczos algorithm \cite{Saad11}. This is an algorithm where only the lowest few eigenvalues and eigenvectors are computed. The elements of the Hamiltonian matrix need not be stored but can be computed only when needed. Using Lanczos algorithm, some studies computed the energy gap \cite{Jorg08,Takahashi10a,Takahashi10b} as well as physical quantities such as the Binder cumulant\cite{Mukherjee15} for up to $N\approx 22$, but further increase in $N$ is hampered by the exponential increase in the dimension of the eigenvectors. The third approach is Quantum Monte Carlo (QMC) \cite{Young08, Young10,Mukherjee15,Ray89}. Unlike the previous methods, instead of including all the terms in Hilbert space, one instead performs a sampling of the states of an effective classical model in imaginary time \cite{Suzuki00}. By combining a novel zero-temperature QMC with a quantum annealing schedule, Das and Chakrabarti computed the exact ground-state of quantum spin-glasses for up to $N=30$ with a numerical accuracy on par with that of Lanczos algorithm\cite{Das08b}. With the traditional QMC, Young et al. are able to study systems with up to 256 spins \cite{Young10}.  

In Lanczos algorithm and QMC, one seeks to obtain the energy levels exactly. The two methods can be viewed as opposite extremes, where the former includes all the terms in Hilbert space while the latter seeks only a representative \emph{random} sample. In this paper, we propose an intermediate between the two. We include only a subspace of the full Hilbert space, but this subspace is constructed not randomly but systematically by including low-lying excitations from the ground-state. Unlike the previous methods, our approach is not exact but an approximation. Nevertheless, intuitively speaking, large excitations should make only minor corrections to the energies of the lowest few energy levels, and might be neglected if some approximation can be tolerated.

The proposal to describe the system using a smaller basis consisting of low-lying excited states is well-grounded on the theoretical framework of Hartree-Fock (HF) approximation in many-body theory \cite{Fetter03} and Configuration-Interaction (CI) in quantum chemistry \cite{Jensen07}. In the HF formalism, one considers a wavefunction consisting of a direct tensor product of single-spin states
\begin{equation}
|0\rangle = |\psi_1\rangle \otimes \cdots \otimes|\psi_N\rangle,
\end{equation}
where $|\psi_i\rangle$ is the state of the $i$th spin. Early on, Lipkin suggested using the HF wavefunction $|0\rangle$ as an approximate ground-state wavefunction for the Heisenberg and Ising spin models with random bonds \cite{Lipkin07}. More recently, Dusuel and Vidal also used $|0\rangle$ in their study of the Lipkin-Meshkov-Glick model \cite{Dusuel05}, a non-disordered system. In particular, the analysis of Dusuel and Vidal showed that the energy obtained from $|0\rangle$ gives just the extensive part of the ground-state energy. Hence, HF approximation by itself is equivalent to mean field approximation. The reason is because the HF wavefunction is a direct product whereas the true ground-state cannot be completely factorized. To improve upon the HF approximation, CI uses $|0\rangle$ as a vacuum state for generating a basis of excited states. For instance, a basis state where spin 1 is excited is 
\begin{equation}
|1\rangle = |\tilde{\psi}_1\rangle \otimes \cdots \otimes|\psi_N\rangle,
\end{equation}
where $|\tilde{\psi}_1\rangle$ arises from exciting spin 1's `ground-state' $|\psi_1\rangle$, and the states of all the other spins remain the same. The ground-state wavefunction is then expanded in terms of a basis consisting of $|0\rangle$ and such excited states. If all possible combinations of exciting all the spins are included the basis is complete and all physical quantities being calculated using CI are exact. In this paper, we make an approximation by including only the one-spin and two-spin excitations.

As mentioned, HF approximation has previously been used by Dusuel and Vidal to study the Lipkin-Meshkov-Glick model \cite{Dusuel05}. Also, the combination of HF approximation and CI is already a standard technique in the field of quantum chemistry \cite{Jensen07}. However, the two methods combined have not yet been applied to the study of quantum spin-glasses before. The purpose of this paper is to perform an extensive numerical study of a specific spin-glass model using a combination of the two methods. Unlike previous approaches based on the full Hilbert space, by focusing on a smaller subspace HF approximation and CI is computationally less expensive, making it possible to study system sizes comparable to that achievable by QMC. Interestingly, even though the dimension of the subspace spanned by these low-lying excitations is very small relative to that of the full Hilbert space, significant improvement in the accuracy of the energies of the levels is achieved by their inclusion. This means that much physical effects are captured by these excitations. Hence, in addition to being an alternative technique for the numerical study of quantum spin-glasses, this approach also provides insights into the relative importance of different parts of Hilbert space spanned by different excitations.

In this paper, we apply these methods to the Sherrington-Kirkpatrick (SK) model in transverse field,
\begin{equation}
H=-\sum_{i=1}^N\sum_{j>i}^NJ_{ij}\sigma_i^{z}\sigma_j^{z} -\Gamma\sum_{i=1}^{N}\sigma_i^x.
\label{eq.II.01}
\end{equation}
$\Gamma$ is the strength of the transverse field, and the couplings $J_{ij}$ are drawn from the gaussian distribution
\begin{equation}
P(J_{ij})=\sqrt{\frac{N}{2\pi J^2}}\exp\left({-\frac{N J_{ij}^2}{2J^2}}\right),
\label{eq.II.02}
\end{equation}
where $J$ measures the strength of the spin-glass term (we set $J=1$ without loss of generality). Several aspects of this model are already well-understood. For instance, its ground-state energy in the thermodynamic limit is known (within the replica-symmetric ansatz) \cite{Thirumalai89}. The absence of replica symmetry breaking in the presence of transverse field has been reported\cite{Ray89}. The phase diagram has also been obtained using both mean field methods \cite{Thirumalai89,Goldschmidt90,Usadel91} and QMC simulations \cite{Mukherjee15,Ray89}. A recent numerical study of its critical behavior reveals the existence of two different universality classes (classical and quantum), with a crossover at a finite temperature \cite{Mukherjee15}. The behavior of the energy gap for small system sizes has also been studied in detail \cite{Takahashi10a,Takahashi10b}. However, the size dependent behavior of the model is still not very clear. In particular, at the point of quantum phase transition, one should observe that the energy gap of an infinite size system becomes zero \cite{Sachdev11}. However, numerical studies of small system sizes ($N\approx 15$) show that the closure of the averaged energy gap does not coincide with the point of phase transition, leading to the proposition that the gap is not a useful quantity to describe phase transition in the SK model \cite{Takahashi10b}. This is an interesting claim but would require numerical evidences from larger system sizes to substantiate. With the approach proposed above, we study larger system sizes and chart the size dependence of the ground-state energy and energy gap. Our results show that for large systems, the behavior of these quantities approaches that predicted by replica theory for an infinite system \cite{Thirumalai89}, and that gap closure is still a good description of phase transition in the SK model.

The paper is divided into two parts. In the first part we study the ground-state energy in detail. For small system sizes (from $N=8$ to 14), we compare the results of HF approximation and CI with the exact results computed using Lanczos algorithm. This allows us to understand the error incurred by our approximations. For larger systems it is no longer possible to obtain the exact ground-state energy, so we compare with the free energy computed using replica theory \cite{Thirumalai89}. A new insight offered by our method is the nature of the leading correction to the extensive part of the ground-state energy when the low-lying excitations are included. The scaling of the correction is `sub-extensive' in the sense that it varies as $\sim N^{0.733}$. This is in stark contrast to the ferromagnetic model Eq. (\ref{eq.I.01}) whose leading correction is known from the Holstein-Primakoff transform to be independent of size \cite{note.intro.01}. 

In the second part, we study the energy gap in detail. Traditionally, the energy gap is obtained by computing the second lowest eigenvalue of the Hamiltonian matrix. Here, we take a slightly novel approach by deriving a formula that relates the energy gap to the ground-state wavefunction. One does not need to solve the eigenvalue problem for the second lowest energy level, making it computationally less expensive. Our gap formula is exact and independent of any approximation scheme. Approximation enters by substituting the HF (or CI) instead of the exact ground-state wavefunction into the formula. For small system sizes, we compare our approximation with the exact gap computed using Lanczos algorithm. When applied to large systems, our method reveals that the energy gap of the SK model decreases as $\sim N^{-0.616}$, much faster than $\sim N^{-0.316}$ for the ferromagnetic model.

The rest of the paper is organized as follows. Secs. \ref{sec.HF approximation}, \ref{sec.CI}, and \ref{sec.truncated CI approx} form the first part of the paper on ground-state energy. Sec. \ref{sec.HF approximation} focuses on HF approximation. We study the transition from the paramagnetic to the ordered phase in finite size systems, the algorithmic aspects of solving the HF stationary point equations, the behavior of the solutions to these equations, and perform benchmark studies of the ground-state energy. Sec. \ref{sec.CI} presents the theoretical formulation of CI within the context of spin systems \cite{note.intro.02}. Sec. \ref{sec.truncated CI approx} presents the benchmark studies of the ground-state energy when low-lying excitations are included using CI. Secs. \ref{sec.gap formula} and \ref{sec.gap.HF approximation} form the second part of the paper on the energy gap. In Sec. \ref{sec.gap formula}, we discuss the complex manner in which the ground-state is promoted into the first excited state in the SK model and derive the gap formula. In Sec. \ref{sec.gap.HF approximation}, the HF and CI wavefunctions are used in the gap formula and the results of numerical studies are presented. Sec. \ref{sec.discussion} discusses and concludes the paper. For the ferromagnetic model Eq. (\ref{eq.I.01}), the HF approximation and CI for both the ground-state energy and energy gap can be computed analytically in the thermodynamic limit. These results, together with those of Holstein-Primakoff transform, are summarized in Appendix \ref{sec.appendix.ferromagnetic model}.

\section{Hartree-Fock approximation}
\label{sec.HF approximation}

The HF wavefunction $|0\rangle$ is defined as the direct product of single-spin states,
\begin{equation}
|0\rangle
=
\prod_{i=1}^{N} 
{\alpha_i \choose \beta_i}
,
\label{eq.III.01}
\end{equation}
where $\alpha_i$ and $\beta_i$ are the spinor components of the $i$th spin in the basis where the Pauli matrix $\sigma_i^{z}$ is diagonal. 
HF approximation uses the variational principle to choose $\alpha_i$ and $\beta_i$ such that the expectation of the Hamiltonian is minimum; i.e., we minimize the function
\begin{equation}
E^{\mathrm{HF}}(\bm{\alpha},\bm{\beta})
=
\langle 0|H|0\rangle
,
\label{eq.III.02}
\end{equation}
with respect to $\bm{\alpha}=(\alpha_1,\cdots,\alpha_N)$ and $\bm{\beta}=(\beta_1,\cdots,\beta_N)$ subjected to the normalization conditions
\begin{equation}
\alpha_i^{2} +\beta_i^{2} =1, \hspace{1.5cm}i=1,\cdots,N.
\label{eq.III.03}
\end{equation}
For the SK model, 
\begin{equation}
E^{\mathrm{HF}}(\bm{\alpha},\bm{\beta})
=
-
\sum_{j>i}
J_{ij}(\alpha_i^2 - \beta_i^2)(\alpha_j^2 - \beta_j^2)
-
2
\Gamma
\sum_{i}
\alpha_i\beta_i
.
\label{eq.III.04}
\end{equation}
Substituting $\beta_i=\sqrt{1-\alpha_i^2}$ and differentiating with respect to $\alpha_i$, the stationary conditions, or HF equations, are 

\begin{equation}
\frac{\partial E^{\mathrm{HF}}}{\partial\alpha_i}
=
\frac{2\Gamma(2\alpha_i^2-1)}{\sqrt{1-\alpha_i^2}}
-
4\alpha_i
\sum_{a\ne i}
J_{ia}(2\alpha_a^2-1)
=0.
\label{eq.III.05}
\end{equation}
When $\Gamma$ is large, the paramagnetic solution
\begin{equation}
\bm{\alpha}_0=\left(\frac{1}{\sqrt{2}},\cdots,\frac{1}{\sqrt{2}}\right),
\label{eq.III.06}
\end{equation}
satisfy all $N$ equations of Eqs. (\ref{eq.III.05}) and is stable. This solution becomes unstable (and lower-energy solutions appear) when the smallest eigenvalue of the Hessian matrix 
\begin{equation}
\left.\frac{\partial E^{\mathrm{HF}}}{\partial\alpha_i\partial\alpha_j}\right|_{\bm{\alpha_0}}
=
8
\left(
\Gamma\delta_{ij}
-
J_{ij}
\right),
\label{eq.III.07}
\end{equation}
becomes negative. $\delta_{ij}$ is the Kronecker delta, $J_{ii}=0$, and $J_{ij}=J_{ji}$. Eq. (\ref{eq.III.07}) is valid for any $N$, lattice geometry, and probability distribution of the bonds $J_{ij}$. For the SK model, we computed $\Gamma_{\mathrm{HF}}$, the $\Gamma$ at which the smallest eigenvalue of the Hessian vanishes, for different realizations of $J_{ij}$ \cite{note01}. The average over the different realizations, $\langle \Gamma_{\mathrm{HF}}\rangle$, as a function of $N$ is shown in Fig. \ref{fig.Fig01}. The HF solutions below $\Gamma_{\mathrm{HF}}$ correspond to symmetry broken states in the ordered phase, and within the HF framework these solutions spontaneously appear at exactly $\Gamma_{\mathrm{HF}}$. Symmetry breaking is, however, not well-defined for small systems. The inset of Fig. \ref{fig.Fig01} shows, for a specific realization of $J_{ij}$ ($N=14$), the energy gaps $E_1-E_0$ (for the paramagnetic regime) and $E_2-E_1$ (for the ordered regime, where $E_0$ and $E_1$ are degenerate) calculated exactly using Lanczos algorithm. We see that it is difficult to pinpoint exactly where the system changes from the paramagnetic to the ordered regime. Nevertheless, let us define the transition into the ordered phase as $\Gamma_{\mathrm{line}}$, the $\Gamma$-intercept of a straight line fitted to $E_1-E_0$ (see inset). The average over different realizations, $\langle\Gamma_{\mathrm{line}}\rangle$, is shown in Fig. \ref{fig.Fig01} \cite{note02}. We see that $\langle \Gamma_{\mathrm{HF}}\rangle$ over-estimates the transition into the ordered phase. As $N\rightarrow\infty$, $\langle \Gamma_{\mathrm{HF}}\rangle\rightarrow 2$, in agreement with replica theory \cite{Thirumalai89}. Analytically, this is also clear because the matrix $J_{ij}$ is drawn from the Gaussian Orthogonal Ensemble and by the semi-circle law its largest eigenvalue $\rightarrow 2$ as $N\rightarrow\infty$.


Below $\Gamma_{\mathrm{HF}}$, we solve Eqs. (\ref{eq.III.05}) numerically with a gradient descent algorithm,
\begin{equation}
\bm{\alpha}^{t+1}
\leftarrow
\bm{\alpha}^{t}
-
\epsilon
\left.\frac{\partial E^{\mathrm{HF}}}{\partial\bm{\alpha}}\right|_{\bm{\alpha}^t}
,
\label{eq.III.08}
\end{equation}
where $\bm{\alpha}^t$ is the estimated solution at step $t$ of iteration, and $\epsilon$ is the step size. Eq. (\ref{eq.III.08}) is iterated until convergence, i.e. $||\bm{\alpha}^{t+1}-\bm{\alpha}^{t}||$ is smaller than some threshold. For a specific realization of $J_{ij}$, the solution for the $\Gamma$ immediately below $\Gamma_{\mathrm{HF}}$ is obtained by first displacing $\bm{\alpha}_0$ slightly along the direction of the eigenvector of the largest eigenvalue of $J_{ij}$ and then iterating Eq. (\ref{eq.III.08}). The solution for the current $\Gamma$ is then used as the initial condition for solving the next lower $\Gamma$. Fig. \ref{fig.Fig03} shows, for $N=10$ and a specific realization of $J_{ij}$, the HF solutions of all 10 spins below $\Gamma_{\mathrm{HF}}$. For comparison, the HF solution for the ferromagnetic model is also shown. 

For small $N$, we can check the quality of HF approximation by comparing $E^{\mathrm{HF}}$ to the exact ground-state energy $E_0$. Since $E^{\mathrm{HF}}$ must necessarily be higher than $E_0$ according to the variational principle, define the excess energy
\begin{equation}
\Delta\mathcal{E}^{\mathrm{HF}}
=
E^{\mathrm{HF}}-E_0,
\end{equation}
for a realization of $J_{ij}$. The average over different realizations, $\langle \Delta \mathcal{E}^{\mathrm{HF}}\rangle$, is shown in Fig. \ref{fig.Fig04} for $N=8$ to 14. HF approximation recovers the exact ground-state energy when $\Gamma=0$ and $\infty$ where the excess energy becomes zero. Although correct at these two extremes, HF approximation over-estimates the ground-state energy as soon as $\Gamma\ne 0$ or $\ne\infty$. $\langle \Delta \mathcal{E}^{\mathrm{HF}}\rangle$ peaks in the region between $\Gamma=1$ and 2, where the system `changes phase' and quantum effects are expected to be strongest.

For large $N$, it is difficult to compute the exact ground-state energy even with Lanczos algorithm due to the large size of the wavefunction. Instead, we compare the HF ground-state energy with the free energy per spin, $\frac{F}{N}$, calculated using replica theory \cite{note05}.  Replica theory is expected to be exact as $N\rightarrow\infty$ . Fig. \ref{fig.Fig06} shows the average HF energy per spin, $\frac{\langle E^{\mathrm{HF}}\rangle}{N}$, for $N=10$ to 1000, and the result of replica theory. It is seen that $\frac{\langle E^{\mathrm{HF}}\rangle}{N}$ approaches $\frac{F}{N}$ for large $N$.

A comment on the accuracy of HF approximation in the limit $\Gamma\rightarrow 0$. The HF energy $E^{\mathrm{HF}}$ reduces to the classical SK Hamiltonian in this limit. This is because when $\Gamma=0$, $\alpha_i$ must be either 0 or 1 in order for the stationary conditions Eq. (\ref{eq.III.05}) to be statisfied. Each $\alpha_i^2-\beta_i^2$ that appears in Eq. (\ref{eq.III.04}) then becomes a binary variable that is either +1 or $-1$. Hence, when $\Gamma=0$, minimizing $E^{\mathrm{HF}}$ is equivalent to minimizing the classical SK energy. This equivalence might be interesting for quantum annealing. If the objective is to arrive at the ground-state configuration of the spin-glass part of the Hamiltonian, then instead of annealing an actual quantum system (such as Eq. (\ref{eq.II.01})), one can instead anneal its HF approximate (i.e., Eq. (\ref{eq.III.04})) which is simpler and might even be implementable in a classical manner. However, whether the HF energy is indeed a more feasible alternative requires further study. 

\section{Configuration Interaction}
\label{sec.CI}

The HF energy over-estimates the true ground-state energy. The HF wavefunction $|0\rangle$ is a direct product, whereas the actual ground-state cannot be completely factorized (unless $\Gamma=0$ or $\infty$). To improve upon the HF approximation, let us first expand the ground-state in a complete basis of $2^N$ linearly independent, direct product states. We create the so-called CI basis from $|0\rangle$ as follows. It is easily shown that $\sigma^y_i$ flips the $i$th spin of $|0\rangle$. Define
\begin{equation}
|i_1\cdots i_k\rangle
=
\sigma_{i_1}^y
\cdots
\sigma_{i_k}^y
|0\rangle,
\label{eq.IV.01}
\end{equation}
where $|i_1\cdots i_k\rangle$ ($i_1<\cdots<i_k$) is a direct product state obtained by flipping the $i_1$th, $\cdots$, and $i_k$th spin of $|0\rangle$. The $\bm{\alpha}$ and $\bm{\beta}$ in $|0\rangle$ in Eq. (\ref{eq.IV.01}) are solutions of the HF equations. The $\sum_{n=0}^N{N\choose n}=2^N$ different ways of flipping provides a complete basis for expanding the trial ground-state wavefunction
\begin{equation}
|\mathrm{CI}\rangle
=
c_0|0\rangle
+
\sum_{\{i\}}c_i|i\rangle
+
\sum_{\{ij\}}c_{ij}|ij\rangle
+
\sum_{\{ijk\}}c_{ijk}|ijk\rangle
+
\cdots,
\label{eq.IV.02}
\end{equation}
where $\sum_{\{i_1\cdots i_k\}}$ denotes summing over ${N\choose k}$ ways to flip $k$ spins and $c_{i_1\cdots i_k}$ are the expansion coefficients. The method of CI uses the variational principle to minimize the CI energy 
\begin{equation}
E^{\mathrm{CI}}
=
\langle \mathrm{CI}|H|\mathrm{CI}\rangle
-
\lambda
\left[
\langle \mathrm{CI}|\mathrm{CI}\rangle-1
\right],
\label{eq.IV.03}
\end{equation}
with respect to the expansion coefficients. $\lambda$ is the Lagrange multiplier to impose the normalization constraint on $|\text{CI}\rangle$. The minimization problem is equivalent to solving the the eigenvalue equation
\begin{equation}
\left(
\begin{array}{cccc}
 H_{00}   &  \cdots  & H_{0a}  & \cdots  \\
 \cdots          &  \cdots  & \cdots         & \cdots  \\
 H_{a0}   &  \cdots  & H_{aa}  & \cdots  \\
 \cdots          &  \cdots  & \cdots         & \cdots  \\
\end{array}
\right)
\left(
\begin{array}{c}
  C_0   \\
  \vdots\\
  C_a\\
  \vdots\\
\end{array}
\right)
=
\lambda
\left(
\begin{array}{c}
  C_0   \\
  \vdots\\
  C_a\\
  \vdots\\
\end{array}
\right),
\label{eq.IV.04}
\end{equation}
where $H_{ab}$ is the ${N\choose a}\times{N\choose b}$  block matrix whose elements are $\langle i_1\cdots i_a|H|j_1\cdots j_b\rangle$ and $C_a$ is a ${N\choose a}$-dimensional column vector whose elements are $c_{i_1\cdots i_a}$. (If $a=0$, then $i_1\cdots i_a$ is 0.) The matrix in Eq. (\ref{eq.IV.04}) is called the CI matrix. The minimum value of $E^{\mathrm{CI}}$ is given by the smallest eigenvalue of the CI matrix and the corresponding eigenvector is the CI ground-state. 

\section{Truncated CI ground-state energy}
\label{sec.truncated CI approx}

Solving Eq. (\ref{eq.IV.04}) with the full CI basis is equivalent to exact diagonalization of the Hamiltonian, but also leads to the same complexity as diagonalizing in any other complete basis set. In the following, let us include only the one-spin and two-spin flips in the expansion Eq. (\ref{eq.IV.02}). Hence, we work with a truncated wavefunction 
\begin{equation}
|\mathrm{CI}^{\prime}\rangle
=
c_0|0\rangle
+
\sum_{i}c_i|i\rangle
+
\sum_{j>i}c_{ij}|ij\rangle
\label{eq.V.00}
\end{equation}
with corresponding CI energy $E^{\mathrm{CI}^{\prime}}=\langle \mathrm{CI}^{\prime}|H|\mathrm{CI}^{\prime}\rangle -\lambda[\langle \mathrm{CI}^{\prime}|\mathrm{CI}^{\prime}\rangle-1]$, and diagonalize a truncated, $\left[1+{N \choose 1}+{N \choose 2}\right]$-dimensional CI matrix. The matrix elements of the truncated CI matrix are given in Appendix \ref{sec.appendix.matrix elements}.

For small $N$, we again compare the truncated CI ground-state energy $E^{\mathrm{CI}^{\prime}}$ with the exact ground-state energy $E_0$. Define the excess energy
\begin{equation}
\Delta \mathcal{E}^{\mathrm{CI}^{\prime}}=E^{\mathrm{CI}^{\prime}} - E_0,
\label{eq.V.01}
\end{equation}
for a single realization of $J_{ij}$. The average over realizations $\langle\Delta \mathcal{E}^{\mathrm{CI}^{\prime}}\rangle$ is shown in Fig. \ref{fig.Fig07} for $N=8$ to 14. To highlight the improvement, the inset shows $\langle\Delta \mathcal{E}^{\mathrm{CI}^{\prime}}\rangle$ and $\langle\Delta \mathcal{E}^{\mathrm{HF}}\rangle$ for $N=14$. Although the one-spin and two-spin flipped states constitute only a small portion of the full basis, they give significant improvements to the ground-state energy, especially in the paramagnetic phase. 

An important difference between SK and the ferromagnetic model is in the correction to the HF ground-state energy $E^{\mathrm{HF}}$ arising from these low-lying excitations. As discussed in Fig. \ref{fig.Fig06}, $E^{\mathrm{HF}}$ scales linearly with $N$ and is an extensive quantity. Let us write the truncated CI matrix as $E^{\mathrm{HF}}\cdot I+R$ where $I$ is the identity matrix and $R$ is the remainder. The smallest eigenvalue of $R$, denoted as $r$, is the correction to $E^{\mathrm{HF}}$. Fig. \ref{fig.Fig08}(a) shows the average over different realizations, $\langle r \rangle$, for $N=8$ to 200. Note the lowering of the curves with increasing $N$. For the ferromagnetic model, shown in the inset, the curves are size-independent for large $N$. The magnitude of $\langle r \rangle$ at the minimum point of each curve (indicated for $N=200$ in Fig. \ref{fig.Fig08}(a)) is plotted against $N$ in Fig. \ref{fig.Fig08}(b). Fitting to the last three points, we see that $-\langle r\rangle_{\mathrm{min.\,pt.}}$ scales as $\sim N^{0.733}$. This means that the ground-state energy of the SK model scales as $aN+bN^{0.733}+\cdots$, whereas that of the ferromagnetic model scales as $a^{\prime}N+b^{\prime}+\cdots$ ($a,b,a^{\prime},$ and $b^{\prime}$ are independent of $N$).

For the ferromagnetic model, the ground-state energy for both HF approximation and truncated CI can be calculated analytically in the limit $N\rightarrow\infty$. These results are summarized in Appendix \ref{sec.appendix.ferromagnetic model.ground-state}.

\section{First excited state and energy gap formula}
\label{sec.gap formula}

The energy gap $\Delta$ is defined as
\begin{equation}
\Delta=E_1-E_0,
\label{eq.VI.01}
\end{equation}
where $E_0$ and $E_1$ are the energies of the ground and first excited states $|E_0\rangle$ and $|E_1\rangle$, respectively. Let us first consider the ferromagnetic model. When $\Gamma^{\prime}=0$, the first excited state is $N$-fold degenerate, but when $\Gamma^{\prime}$ is turned on slightly the degeneracy is lifted and the state
\begin{equation}
\frac{1}{\sqrt{N}}\sum_{i=1}^{N}|i\rangle
\label{eq.VI.02}
\end{equation}
splits away to become the first excited state. Note that Eq. (\ref{eq.VI.02}) is a superposition of states $|i\rangle$ which are only one-spin flip from the HF ground-state $|0\rangle$. The situation is slightly more complicated for the SK model. We computed the exact ground-state and first excited state spin configurations at $\Gamma=0$ for different realizations of $J_{ij}$. For a specific realization, let $\nu$ denote the number of spins in the configuration of the ground-state which are flipped in the first excited state. Fig. \ref{fig.Fig09} shows the distribution of $\nu$ computed from 10000 realizations of $J_{ij}$ ($N=12$). Although $\nu=1$ for $>50\%$ of the realizations, a significant proportion ($\approx 40\%$) has $\nu\ge 2$.

Let us consider a Hermitian operator $A$ that flips the appropriate spin(s) in the ground-state to generate the first excited state, i.e.,
\begin{equation}
|E_1\rangle=A|E_0\rangle.
\label{eq.VI.03}
\end{equation}
Define the generating function
\begin{equation}
G(\gamma)=\langle E_0  |e^{-i\gamma A}He^{i\gamma A}|E_0  \rangle,
\label{eq.VI.04}
\end{equation}
where $\gamma$ is a parameter. Expanding the right side of Eq. (\ref{eq.VI.04}) to second order in $\gamma$, we have
\begin{equation}
\langle E_0  |e^{-i\gamma A}He^{i\gamma A}|E_0  \rangle
=
\langle E_0|
\left[
H
+
i\gamma
[H,A]
-
\frac{\gamma^2}{2}
(HA^2+A^2H-2AHA)
+O(\gamma^3)
\right]
|E_0\rangle.
\label{eq.VI.05}
\end{equation}
On the left side, the Taylor expansion of $G(\gamma)$ is $G(0)+\frac{\partial G}{\partial \gamma}|_{\gamma=0}\gamma+\frac{1}{2}\frac{\partial^2 G}{\partial \gamma^2}|_{\gamma=0}\gamma^2+O(\gamma^3)$. Equating the $\gamma^2$ terms, we arrive at the gap formula
\begin{equation}
\Delta
=
\frac{1}{2}
\frac{1}{\langle E_0|A^2|E_0\rangle}
\left.\frac{\partial^2 G}{\partial \gamma^2}\right|_{\gamma=0},
\label{eq.VI.06}
\end{equation}
where we have used $\langle E_0|HA^2+A^2H|E_0\rangle=2E_0\langle E_0|A^2|E_0\rangle$, $E_1=\langle E_1|H|E_1\rangle / \langle E_1|E_1\rangle$, and Eq. (\ref{eq.VI.01}).  

The gap formula Eq. (\ref{eq.VI.06}) is an exact relation that depends only on the exact ground-state $|E_0\rangle$ and an appropriate $A$. However, it is in general difficult to obtain the exact ground-state, especially for large $N$. Also, as discussed above, for the SK model the choice of $A$ depends on the realization of $J_{ij}$ since $\nu$ may be different for different $J_{ij}$. To overcome the first difficulty, we propose making an approximation by replacing $|E_0\rangle$ with the HF ground-state $|0\rangle$ or the truncated CI ground-state $|\mathrm{CI}^{\prime}\rangle$. Concerning the latter difficulty, as we have seen that $\nu=1$ is the most common case, in the following we shall restrict ourselves to an $A$ that makes one-spin flips to the ground-state.

\section{Energy gap from one-spin flip first excited state}
\label{sec.gap.HF approximation}

\subsection{One-spin flip operator}
We define the one-spin flip operator  
\begin{equation}
A_1=\sum_{i=1}^{N} y_i\sigma_i^y.
\label{eq.VI.07}
\end{equation}
The Pauli matrix $\sigma_i^y$ flips the $i$th spin and the real parameter $y_i$ describes the contribution of the flipped spin to the first excited state. The $y_i$s are constrained by the condition
\begin{equation}
\langle E_0|(A_1)^2| E_0 \rangle=1,
\label{eq.VI.08}
\end{equation}
which normalizes the excited state generated by $A_1$. The excited state Eq. (\ref{eq.VI.02}) of the ferromagnetic model is a special case of $A_1|E_0\rangle$ with $|E_0\rangle=|0\rangle$ and $y_i=1/\sqrt{N}$. For the SK model, the parameters $y_i$ are not known \emph{a priori} and depend on the realization of $J_{ij}$. Replacing $A$ in Eq. (\ref{eq.VI.06}) by $A_1$, the energy gap becomes a function of $y_i$. The gap is minimized with respect to $y_i$ to determine which spins are flipped in the first excited state.

\subsection{HF approximation}

We first consider the HF approximation of Eq. (\ref{eq.VI.06}). Replacing $A$ and $|E_0\rangle$ in Eq. (\ref{eq.VI.04}) by $A_1$ and $|0\rangle$, we have  
\begin{equation}
G_1^{\mathrm{HF}}(\gamma)=\langle 0|e^{-i\gamma A_1}He^{i\gamma A_1}|0\rangle.
\label{eq.VII.01}
\end{equation}
Replacing $G$, $A$, and $|E_0\rangle$ on the right side of Eq. (\ref{eq.VI.06}) by $G_1^{\mathrm{HF}}$, $A_1$, and $|0\rangle$, we have the HF energy gap
\begin{equation}
\Delta_1^{\mathrm{HF}}
=
-8
\sum_{i}
\sum_{j\ne i}J_{ij}\alpha_i\alpha_j\beta_i\beta_j y_i y_j
+
\Gamma
\sum_{i}\frac{y_i^2}{\alpha_i\beta_i},
\label{eq.VII.02}
\end{equation}
where $\bm{\alpha},\bm{\beta}$ are solutions of the HF equations, and the subscript 1 in $G_1^{\mathrm{HF}}$ and $\Delta_1^{\mathrm{HF}}$ serves to remind us that $A_1$ is used in place of $A$. The constraint Eq. (\ref{eq.VI.08}) becomes
\begin{equation}
\sum_{i=1}^N y_i^2=1.
\label{eq.VII.03}
\end{equation}
The derivation of Eq. (\ref{eq.VII.02}) is given in Appendix \ref{sec.appendix.derivation HF gap}.

$\Delta_1^{\mathrm{HF}}$ is a quadratic form of $y_i$ and is easily minimized subjected to the condition Eq. (\ref{eq.VII.03}). Fig. \ref{fig.Fig10}(a) shows $\Delta_1^{\mathrm{HF}}$ for a realization of $J_{ij}$ with $\nu=1$ ($N=8$). In the ordered regime ($\Gamma<0.5$), $\Delta_1^{\mathrm{HF}}$ agrees very well with the exact gap $E_2-E_1$ \cite{note05.5}. In the paramagnetic regime ($\Gamma>\Gamma_{\mathrm{HF}}\approx1.5$), $\Delta_1^{\mathrm{HF}}$ is displaced below the curve $E_1-E_0$. The inset shows the same quantities for another realization of $J_{ij}$ with $\nu=2$. The results are similar except that $\Delta_1^{\mathrm{HF}}$ does not approach $E_2-E_1$ as $\Gamma\rightarrow 0$. In this case, the disagreement in the ordered regime is to be expected as one should use an $A$ that makes two-spin flips instead of one-spin flip. 

Fig. \ref{fig.Fig12} shows the solutions of $y_i$ for the realization with $\nu=1$ in Fig. (\ref{fig.Fig10})(a). In the ordered regime, there are two large components (red, dashed lines) and the first excited state is a superposition of mainly these two components. In the paramagnetic regime, $\Delta_1^{\mathrm{HF}}$ is minimized by the eigenvector of the largest eigenvalue of $J_{ij}$ and is independent of $\Gamma$.   

Fig. \ref{fig.Fig10}(b) shows $\langle \Delta_1^{\mathrm{HF}}\rangle$, $\langle E_1-E_0\rangle$, and $\langle E_2-E_1\rangle$, the average of $\Delta_1^{\mathrm{HF}}$, $E_1-E_0$, and $E_2-E_1$ over different realizations of $J_{ij}$ with $\nu=1$ ($N=14$) \cite{note06}. The results are similar to that of a single realization. The inset highlights the region in the ordered regime and shows that the average absolute error $\langle|\Delta_1^{\mathrm{HF}}-(E_2-E_1)|\rangle$ and its fluctuation indeed vanish as $\Gamma\rightarrow 0$.

For large $N$, it is no longer possible to compute the energy gap exactly for comparison. Furthermore, the $\nu$ of a particular realization of $J_{ij}$ is also unknown. Nevertheless, we apply Eq. (\ref{eq.VII.02}) to all the realizations of $J_{ij}$ that we sampled and the average gap $\langle \Delta_1^{\mathrm{HF}}\rangle$ is computed by summing over all realizations regardless of whether $\nu=1$ or not. In the ordered phase, the average gap computed in this way is therefore an overestimation of the actual gap. This is because applying $A_1$ to a realization with $\nu>1$ necessarily promotes the ground-state to a higher state than the first excited state \cite{note07}. Fig. \ref{fig.Fig11} shows $\langle \Delta_1^{\mathrm{HF}}\rangle$ from $N=10$ to 1000. As $N$ increases, the minimum of the energy gap (indicated for the curve of $N=20$ in the figure) approaches asymptotically towards $\Gamma=2$, the point of phase transition. Hence, at least within the HF framework, we verified that at a quantum phase transition the energy gap goes to zero \cite{Sachdev11}.

\begin{figure}[h]
\begin{center}
\includegraphics[scale=0.75]{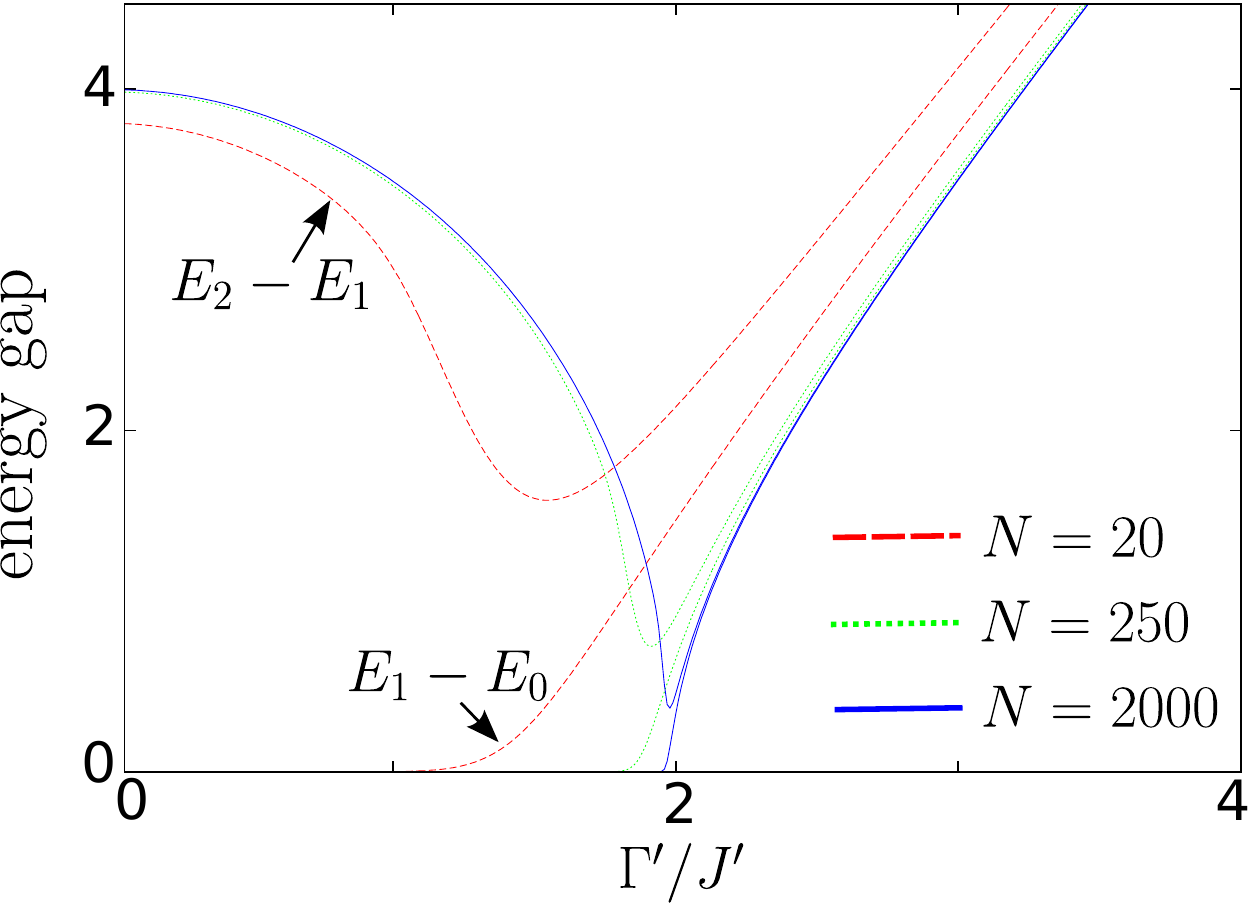}
\caption{(Color online) Energy gap of the infinite range ferromagnetic model in transverse field for various $N$. For each $N$, the bottom curve is the energy difference between the ground and first excited states $E_1-E_0$, and the top curve is for that between the first and second excited states $E_2-E_1$.}
\label{fig.Fig00}
\end{center}
\end{figure}

\begin{figure}[h]
\begin{center}
\includegraphics[scale=0.75]{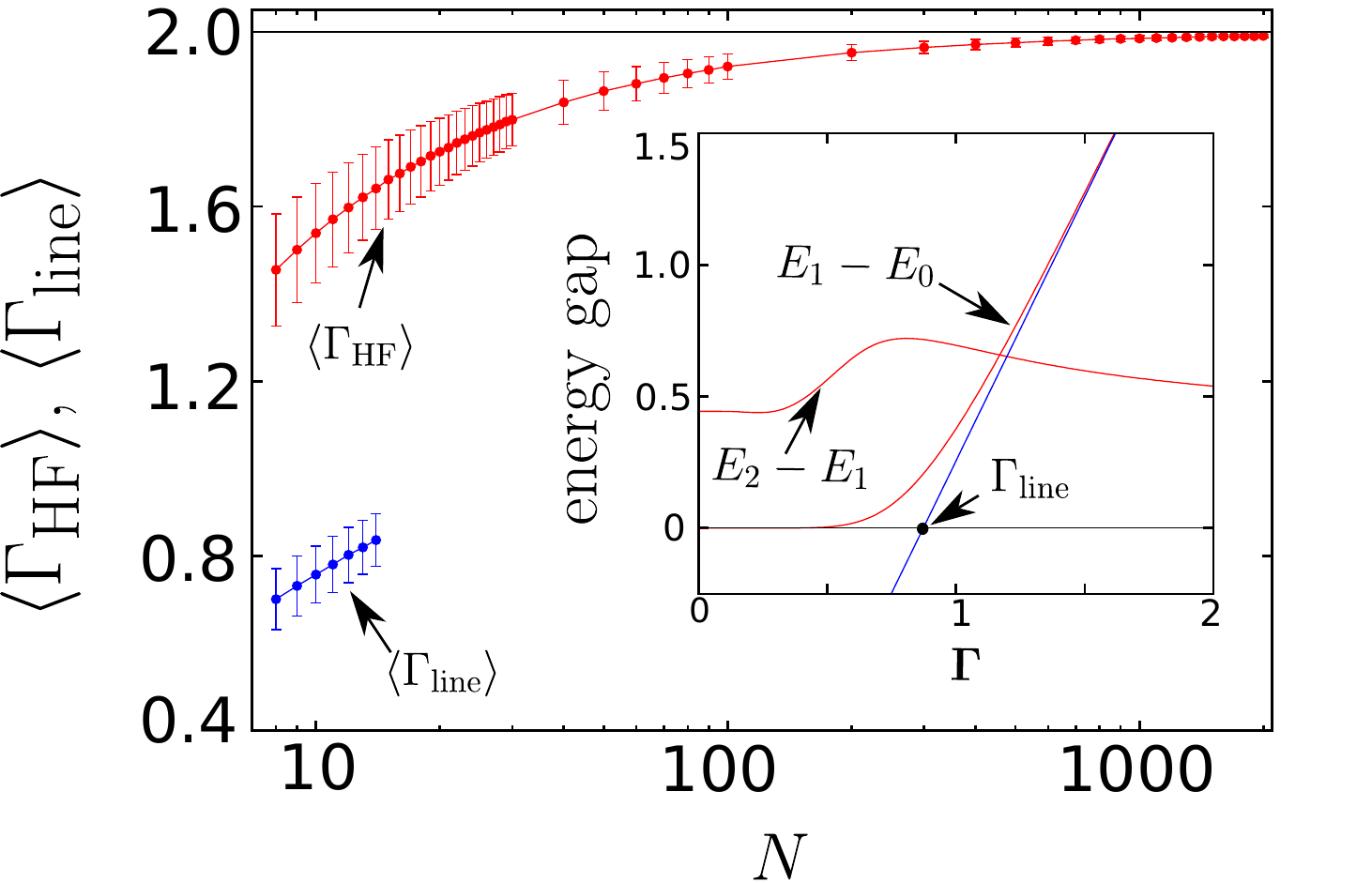}
\caption{(Color online) $\langle \Gamma_{\mathrm{HF}}\rangle$ ($\langle \Gamma_{\mathrm{line}}\rangle$), the average of $\Gamma_{\mathrm{HF}}$ ($\Gamma_{\mathrm{line}}$) over different realizations of $J_{ij}$, for various $N$. Error bars indicate standard deviation. Connecting lines are to guide the eye only. Inset: Example of the exact gaps $E_1-E_0$ and $E_2-E_1$ for a specific realization of $J_{ij}$ ($N=14$). $\Gamma_{\mathrm{line}}$ is defined as the $\Gamma$-intercept of the straight line fitted to $E_1-E_0$, as shown.}
\label{fig.Fig01}
\end{center}
\end{figure}

\begin{figure}[h]
\begin{center}
\includegraphics[scale=0.75]{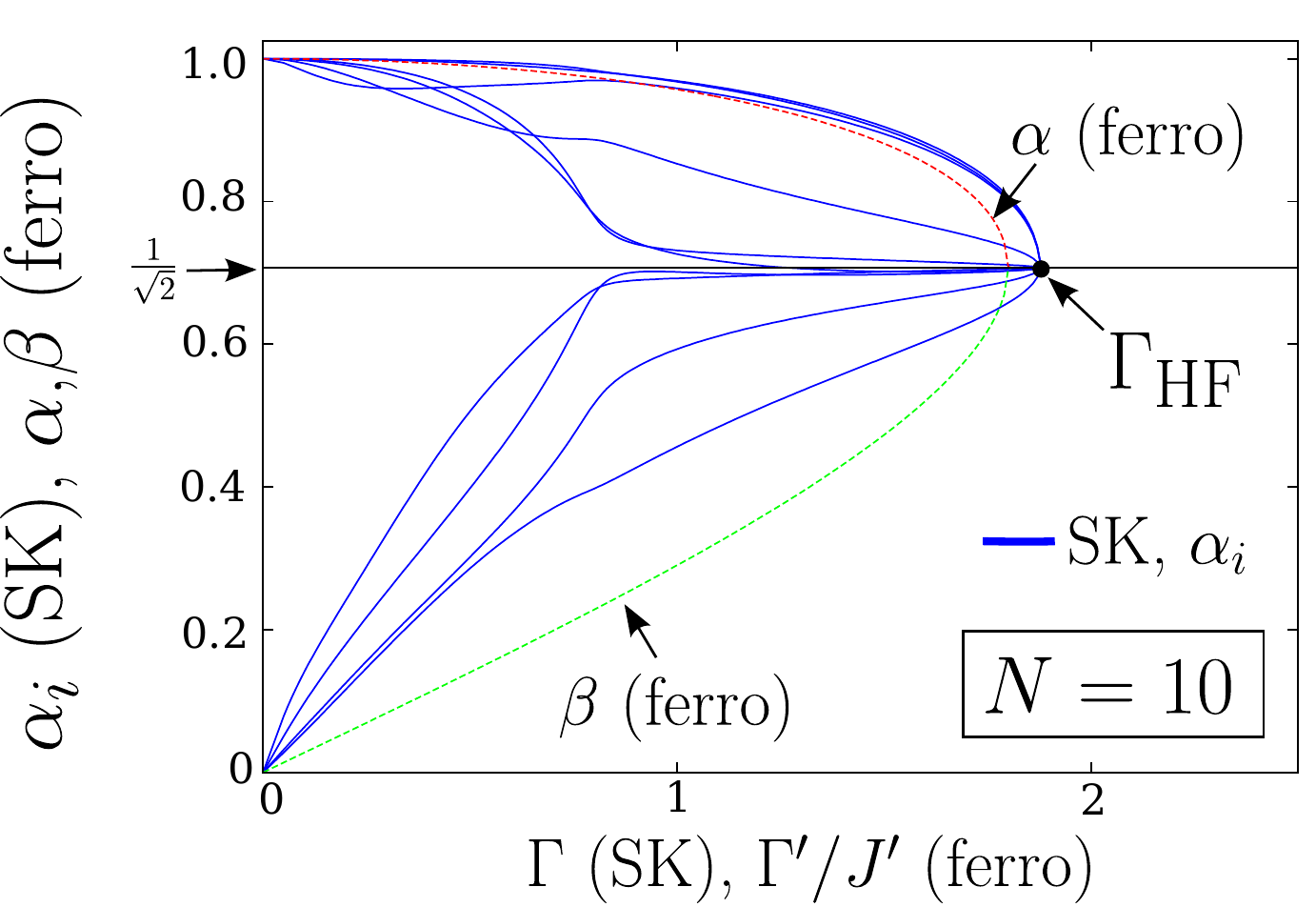}
\caption{(Color online) Solid lines: HF solutions below $\Gamma_{\mathrm{HF}}$ for a specific realization of $J_{ij}$ ($N=10$). Dashed lines: HF solution for the ferromagnetic model of the same $N$.}
\label{fig.Fig03}
\end{center}
\end{figure}

\begin{figure}[h]
\begin{center}
\includegraphics[scale=0.75]{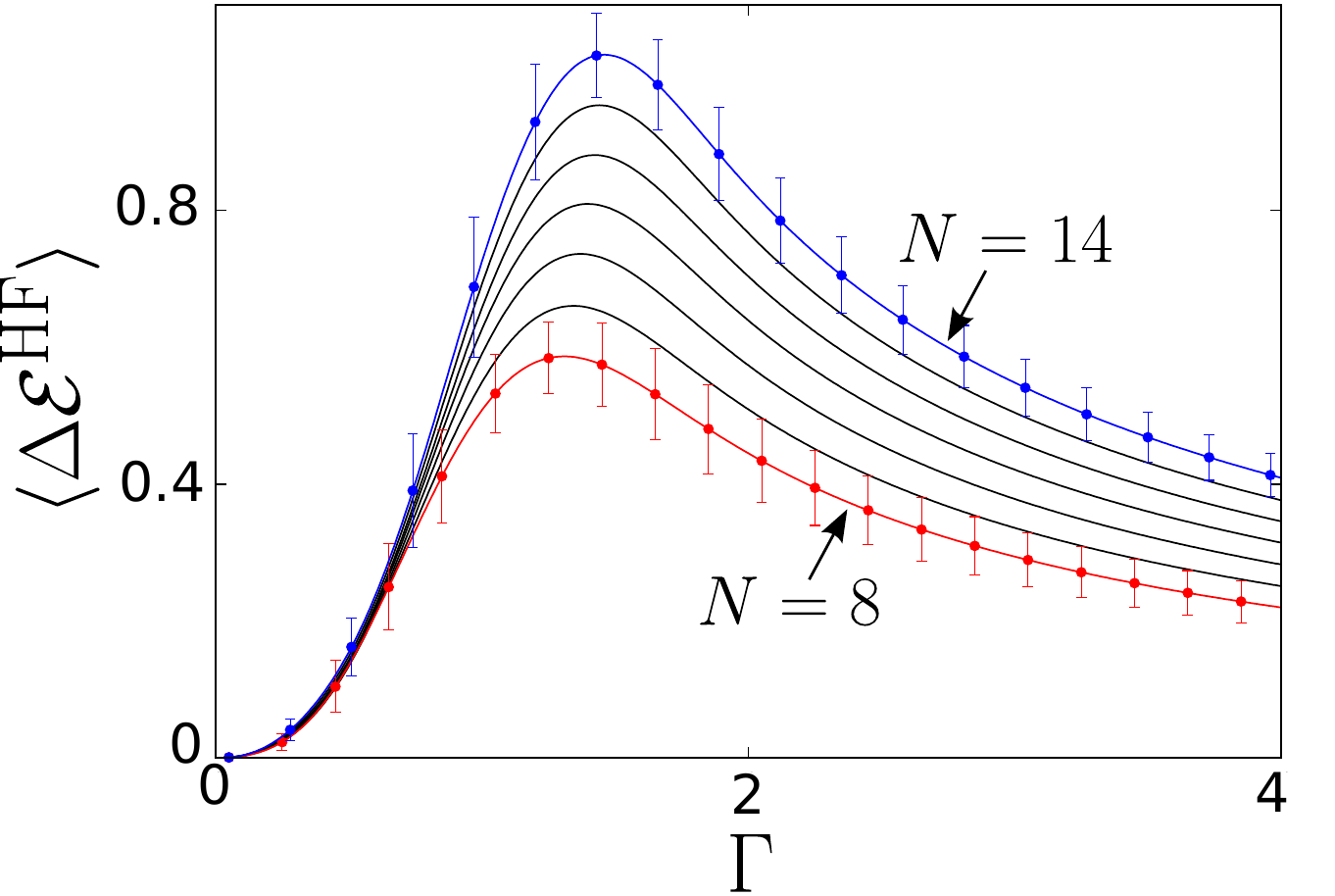}
\caption{(Color online) Average excess energy $\langle \Delta \mathcal{E}^{\mathrm{HF}}\rangle$ for $N=8$ to 14. Error bars for $N=8$ and 14 indicate standard deviation.}
\label{fig.Fig04}
\end{center}
\end{figure}

\begin{figure}[h]
\begin{center}
\includegraphics[scale=0.75]{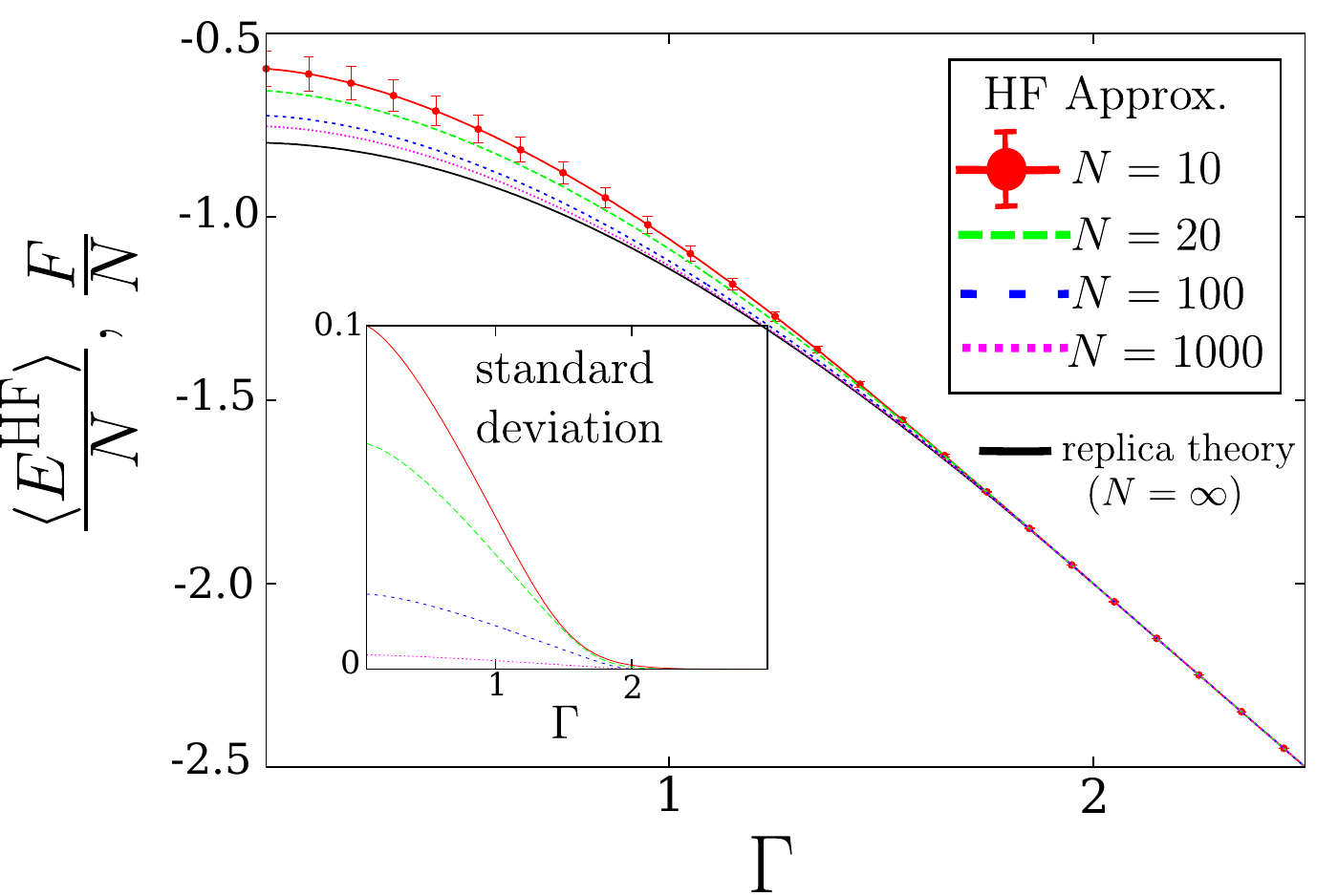}
\caption{(Color online) Average HF ground-state energy per spin, $\frac{\langle E^{\mathrm{HF}} \rangle}{N}$, for various $N$, and free energy per spin, $\frac{F}{N}$, computed using the replica-symmetric mean-field equations ($\beta\rightarrow\infty$) of Ref. \cite{Thirumalai89}. Error bars for $N=10$ indicate standard deviation. Standard deviations for individual $N$ are also shown in the inset using the same line type.}
\label{fig.Fig06}
\end{center}
\end{figure}

\begin{figure}[h]
\begin{center}
\includegraphics[scale=0.75]{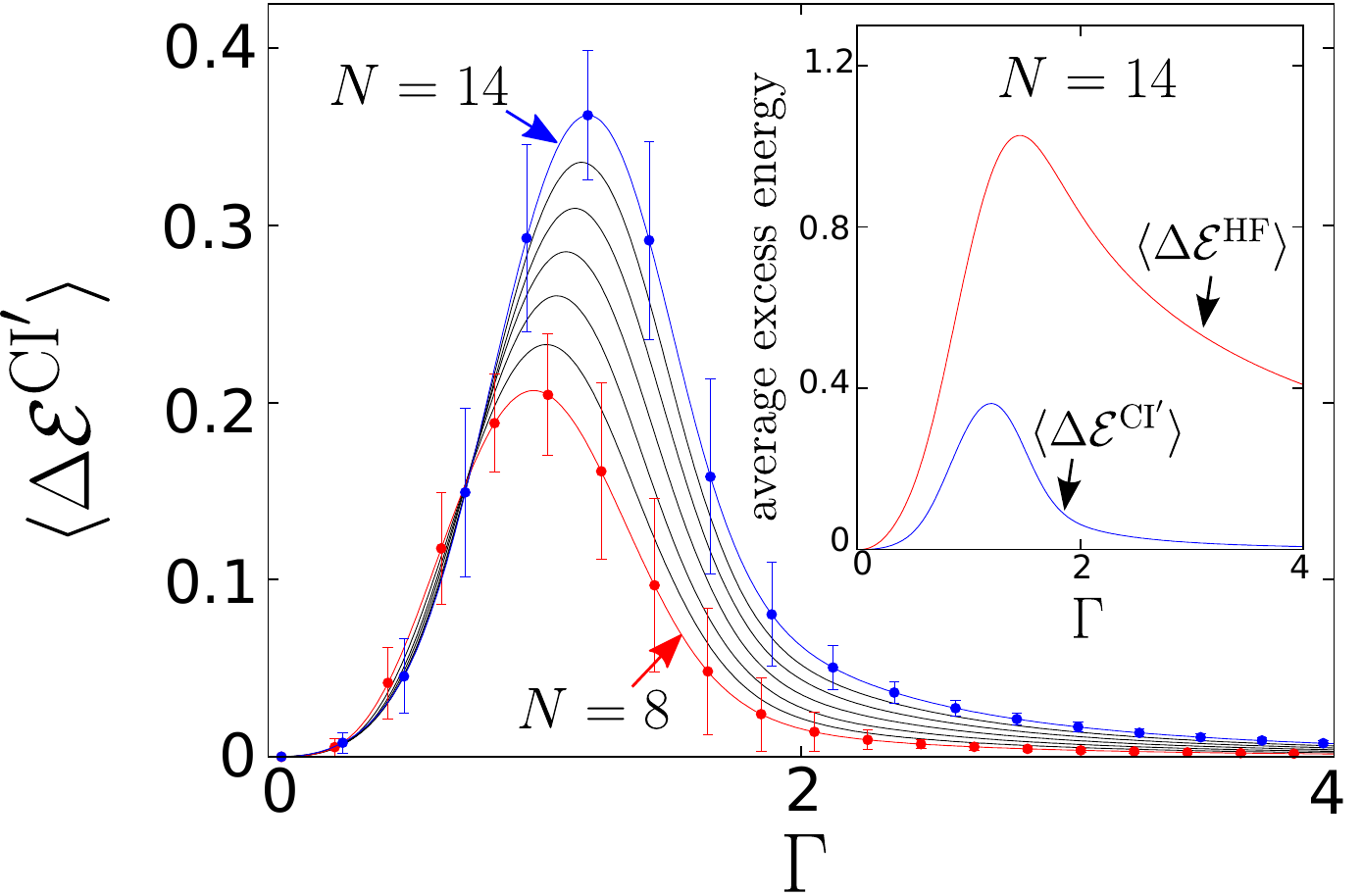}
\caption{(Color online) Average excess energy $\langle \Delta \mathcal{E}^{\mathrm{CI}^{\prime}} \rangle$ for $N=8$ to 14. Error bars for $N=8$ and 14 indicate standard deviation. Inset: Comparing $\langle \Delta\mathcal{E}^{\mathrm{HF}} \rangle$ and $\langle \Delta \mathcal{E}^{\mathrm{CI}^{\prime}} \rangle$ for $N=14$.}
\label{fig.Fig07}
\end{center}
\end{figure}

\begin{figure}[h]
\begin{center}
\includegraphics[scale=0.75]{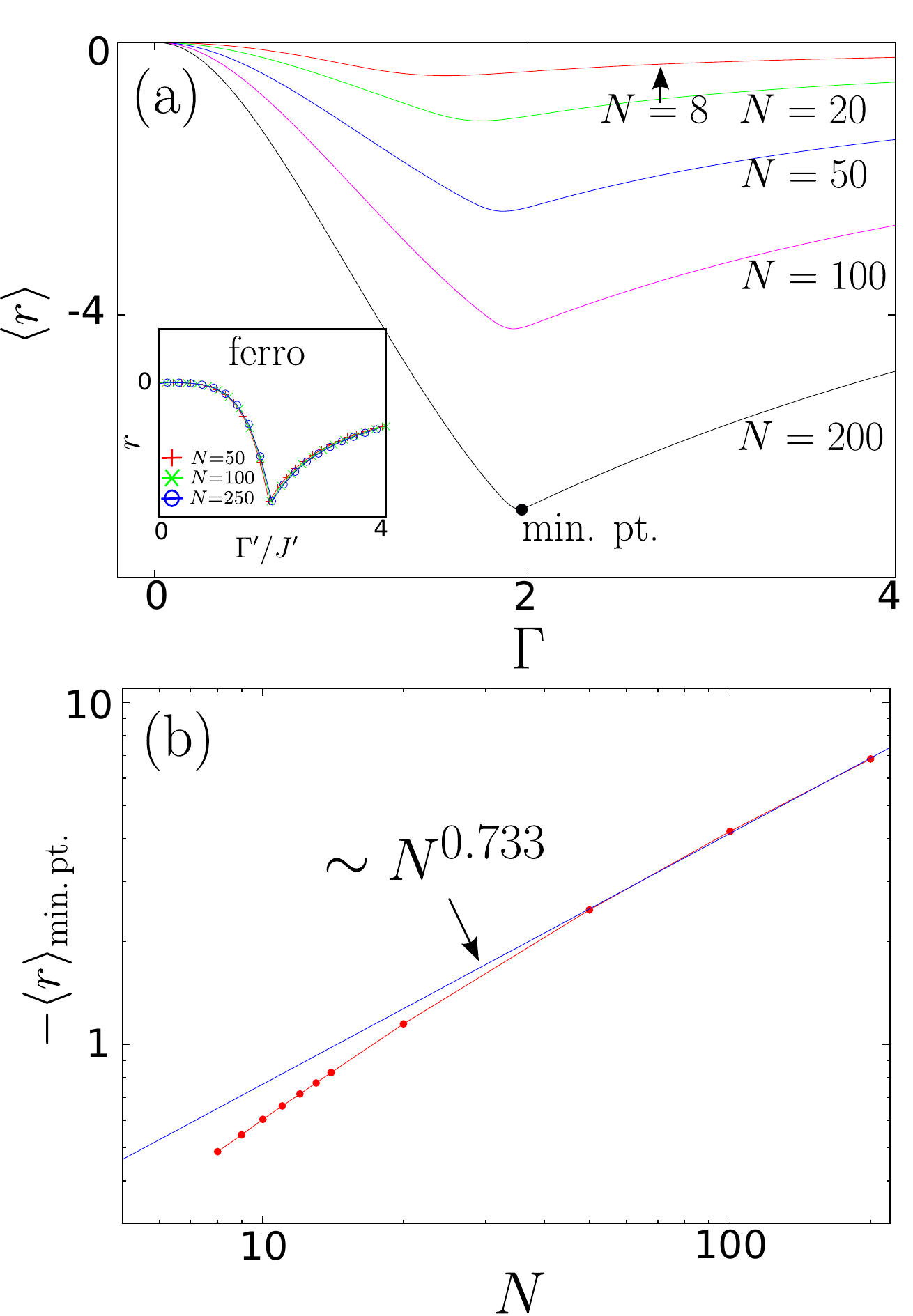}
\caption{(Color online) (a) $\langle r\rangle$, the average over different realizations of the smallest eigenvalue of $R$ for the SK model. Inset: Eigenvalue $r$ of $R$ for the ferromagnetic model. (b) Scaling of $-\langle r\rangle_{\mathrm{min.\,pt.}}$ with $N$ for the SK model. The magnitude of the value of $\langle r\rangle$ at the minimum point of each curve (c.f. panel (a)) is plotted against $N$. Lines (red) connecting the dots are to guide the eye only. Upper (blue) line is a straight line fitted to the points for $N=50$, 100, and 200.}
\label{fig.Fig08}
\end{center}
\end{figure}

\begin{figure}[h]
\begin{center}
\includegraphics[scale=0.75]{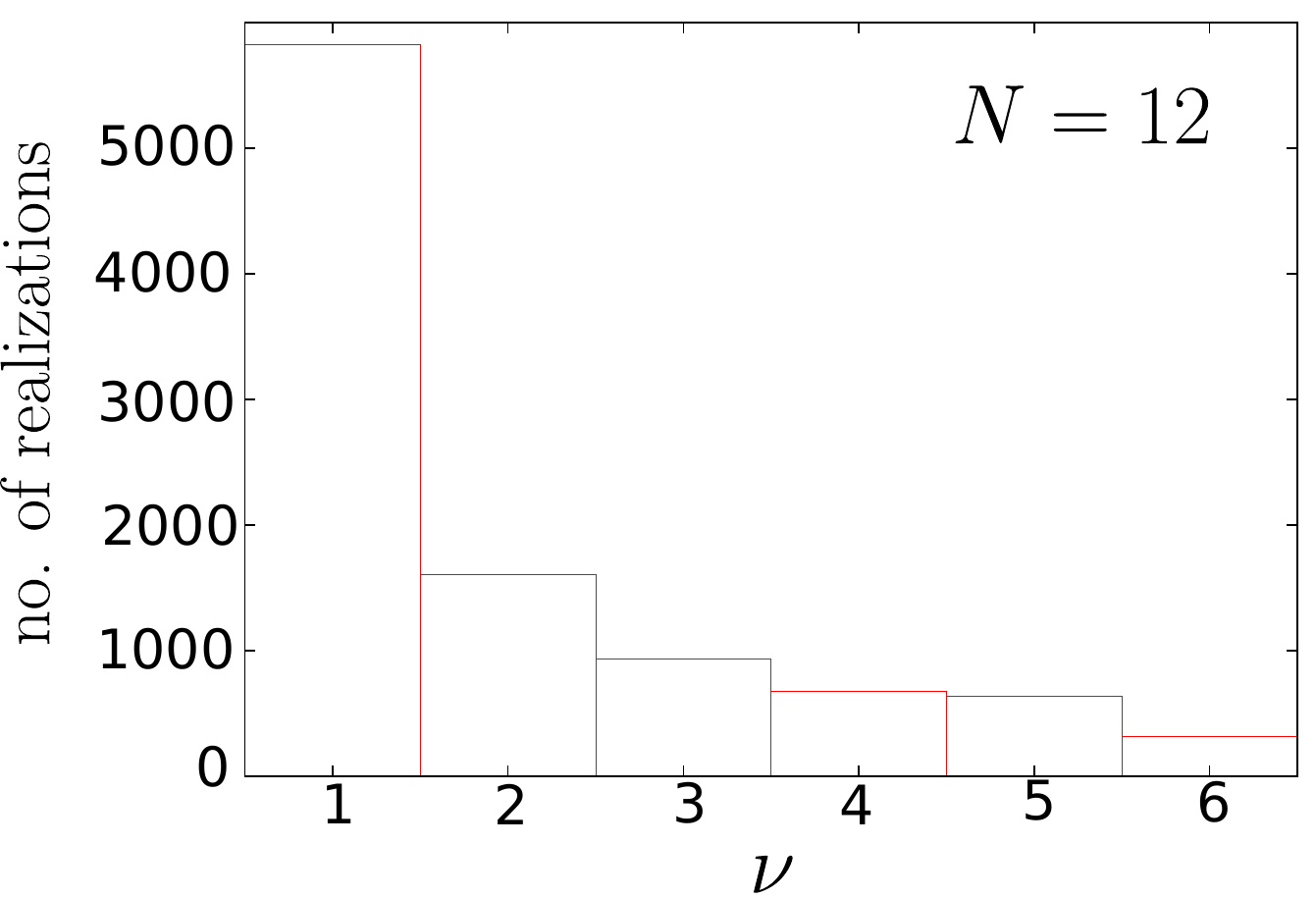}
\caption{(Color online) Distribution of $\nu$, the number of spins in the ground-state configuration which are flipped in the first excited state, for 10000 different realizations of $J_{ij}$. Results are for $N=12$, and the spin configurations are calculated at $\Gamma=0$. }
\label{fig.Fig09}
\end{center}
\end{figure}

\begin{figure}[h]
\begin{center}
\includegraphics[scale=0.75]{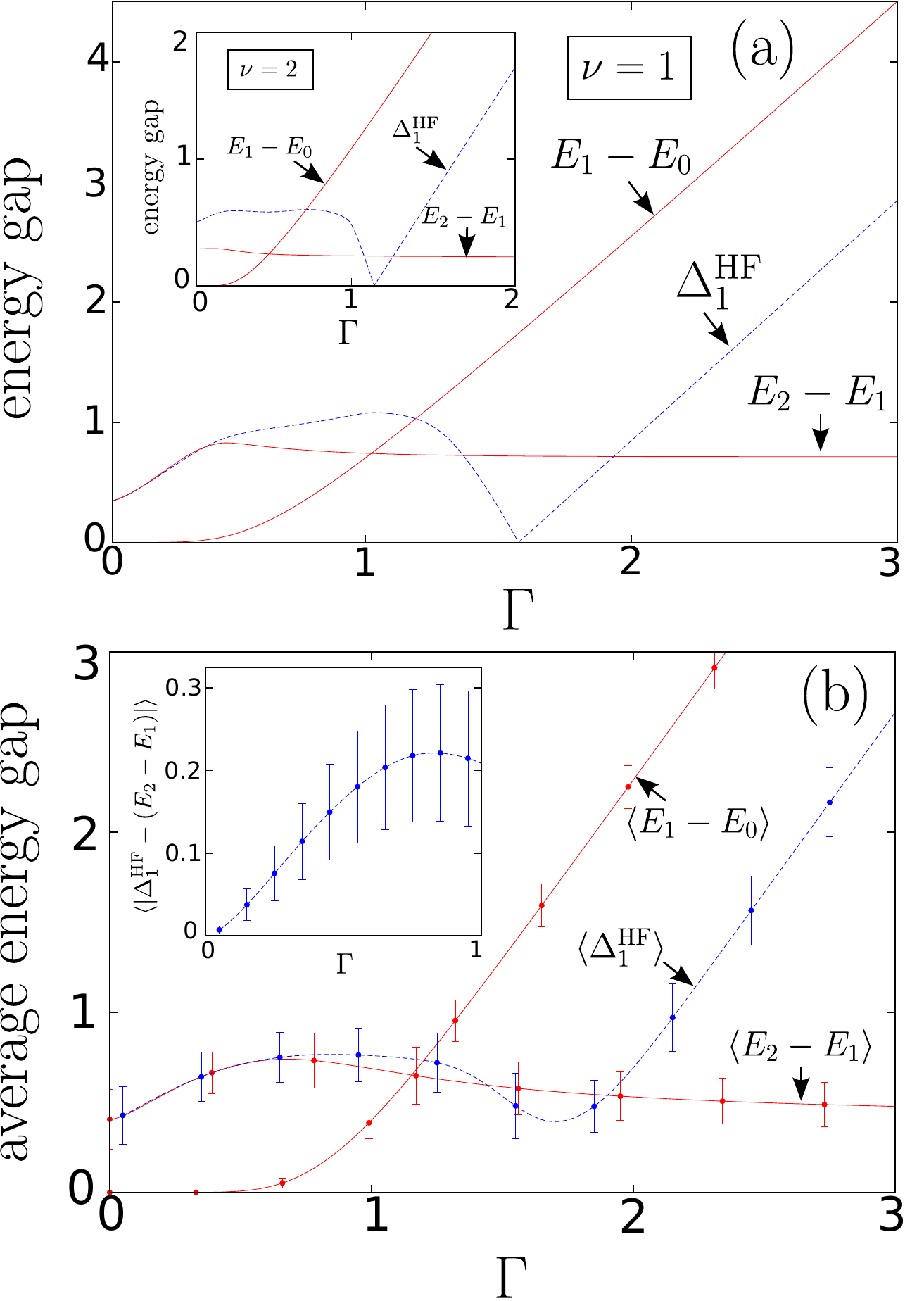}
\caption{(Color online) (a) Energy gaps $\Delta_1^{\mathrm{HF}}$ (blue, dashed line), $E_1-E_0$, and $E_2-E_1$ (red, solid lines) for a single realization of $J_{ij}$ with $\nu=1$ ($N=8$). Inset: Same quantities for a realization with $\nu=2$. (b) Average energy gaps $\langle\Delta_1^{\mathrm{HF}}\rangle$ (blue, dashed line), $\langle E_1-E_0\rangle$, and $\langle E_2-E_1\rangle$ (red, solid lines) taken over different realizations of $J_{ij}$ with $\nu=1$ ($N=14$). Error bars indicate standard deviation. Inset: Average absolute error of the gap, $\langle |\Delta_1^{\mathrm{HF}}-(E_2-E_1)| \rangle$, in the ordered regime. Error bars indicate standard deviation.}
\label{fig.Fig10}
\end{center}
\end{figure}

\begin{figure}[h]
\begin{center}
\includegraphics[scale=0.75]{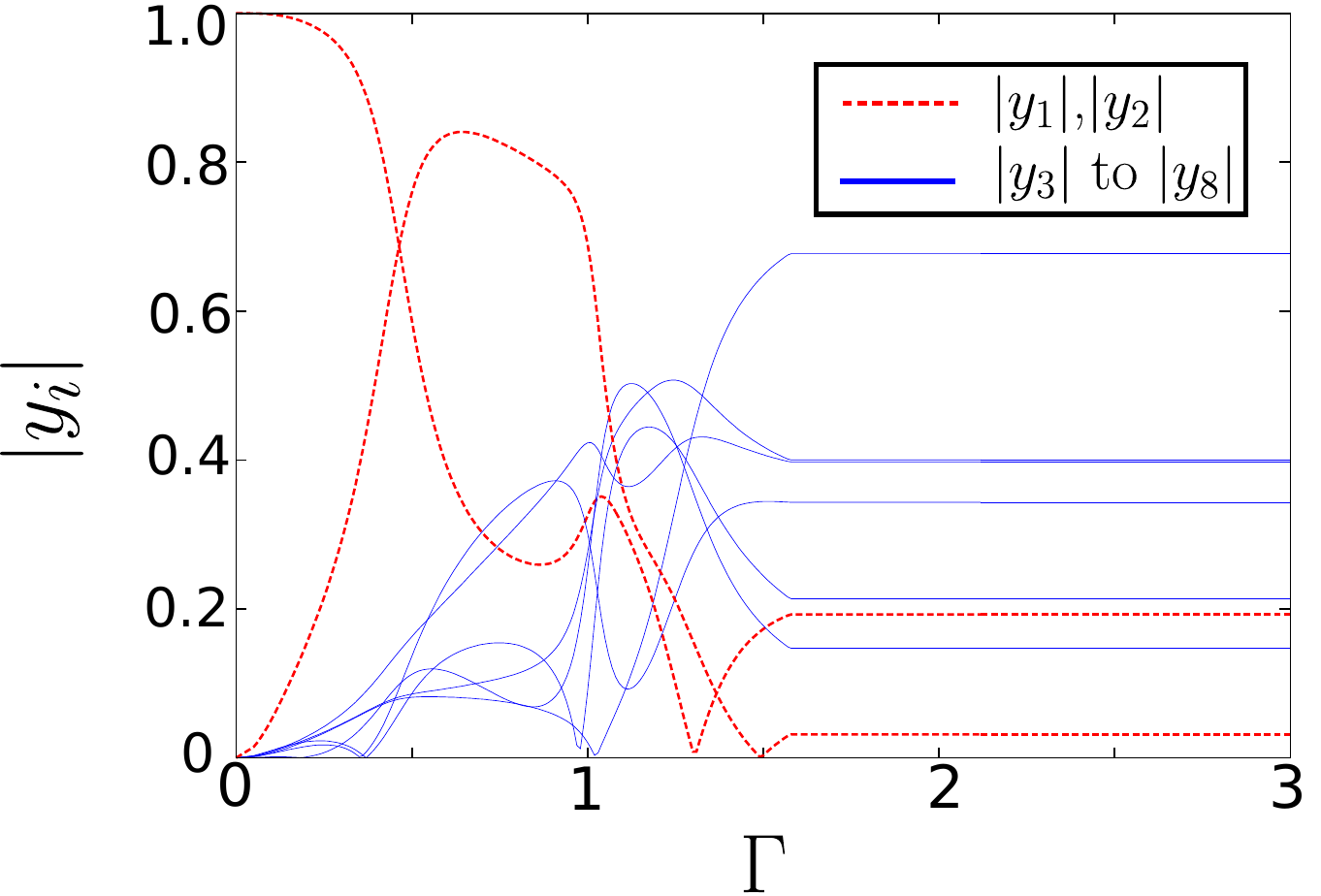}
\caption{(Color online) Solutions of $y_i$ that minimize Eq. (\ref{eq.VII.02}) for the realization of $J_{ij}$ shown in Fig. \ref{fig.Fig10}(a) ($\nu=1$). The absolute value of each component $|y_i|$ is plotted against $\Gamma$. In the ordered regime ($\Gamma<0.5$), two of the components, $|y_1|$ and $|y_2|$ (red, dashed lines), contribute much more than the rest (blue, solid lines).}
\label{fig.Fig12}
\end{center}
\end{figure}

\begin{figure}[h]
\begin{center}
\includegraphics[scale=0.75]{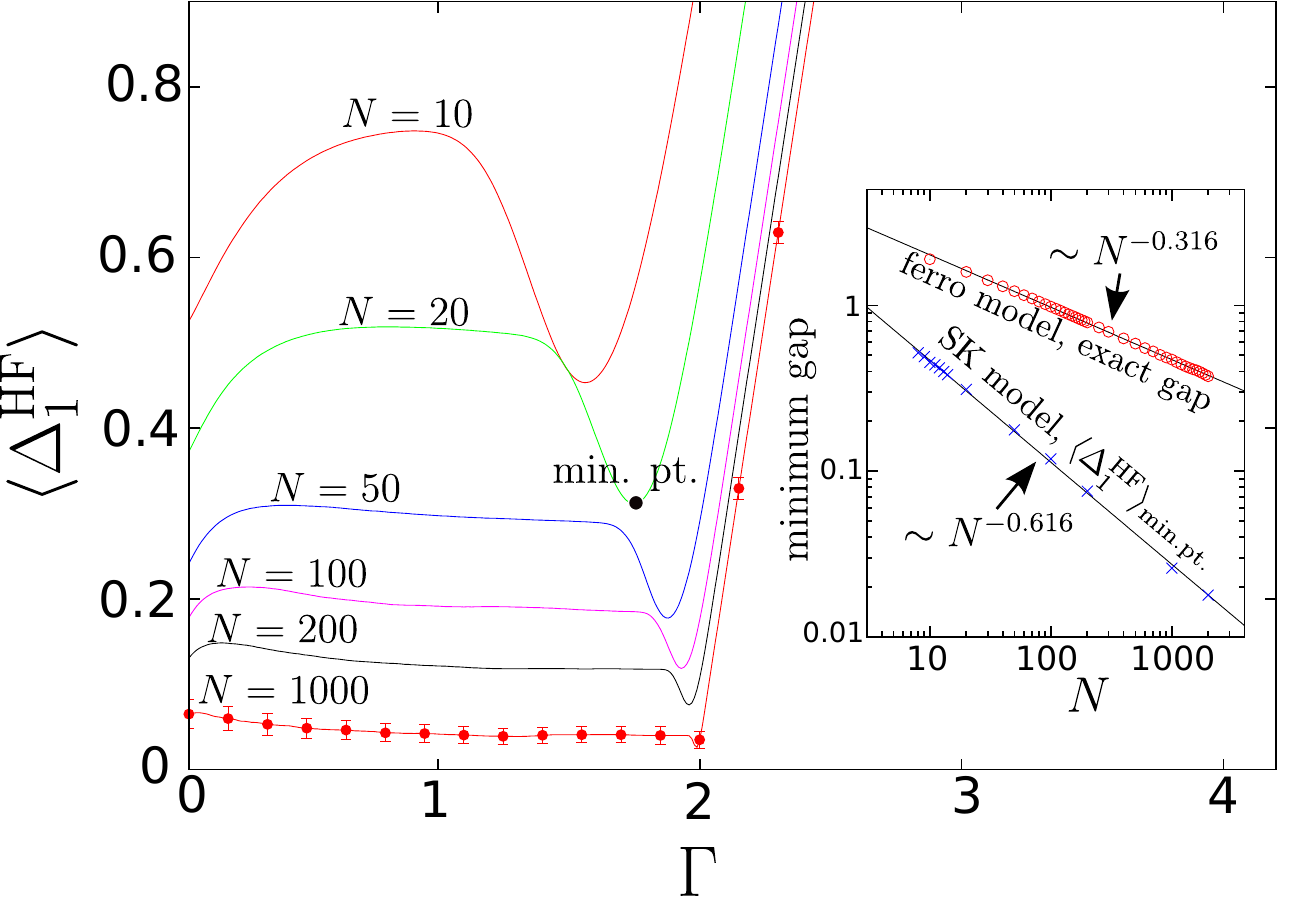}
\caption{(Color online) $\langle \Delta_1^{\mathrm{HF}}\rangle$, the average of the HF energy gap over different realizations of $J_{ij}$ for various $N$. Error bars for $N=1000$ indicate standard deviation.  Inset: Scaling of the minimum energy gap with $N$ for the SK model (crosses, blue) and the ferromagnetic model (circles, red). The solid lines are straight lines fitted to the data points.}
\label{fig.Fig11}
\end{center}
\end{figure}

It is interesting to compare the energy gap between the SK model and the ferromagnetic model (Fig. \ref{fig.Fig00}). Firstly, the gap of the SK model in the ordered phase decreases to zero as $N\rightarrow\infty$, whereas that of the ferromagnetic model remains finite. From previous studies, the classical SK model (i.e., $\Gamma=0$) is already known to have many energetically degenerate ground-states in the thermodynamic limit. Here, we observe numerically that when $\Gamma>0$, the energies of the two lowest levels remain very close to each other in the ordered phase all the way till the critical point. Hence, unlike the ferromagnetic model, the ground-state and the first excited state of the SK model are not well separated in energy and this might present difficulties for the quantum annealing of disordered spin models exhibiting \emph{continuous} transitions (to be discussed in Sec. \ref{sec.discussion}). Secondly, as $N$ increases the minimum energy gap of the SK model decreases much faster than that of the ferromagnetic model. For the former, the minimum gap is defined as the minimum of the $\langle \Delta_1^{\mathrm{HF}}\rangle$ curve (c.f. $N=20$ curve in Fig. \ref{fig.Fig11}). For the latter, it is defined as the minimum of the $E_2-E_1$ (i.e. top) curve shown in Fig. \ref{fig.Fig00}. The inset of Fig. \ref{fig.Fig11} shows that the minimum gap of the SK and the ferromagnetic model scale as $\sim N^{-0.616}$ and $\sim N^{-0.316}$, respectively. Hence, the SK model is much more difficult to anneal across the critical point than the ferromagnetic model.

\subsection{Truncated CI}

We now improve upon the HF approximation by using the truncated CI wavefunction as the ground-state. The generating function is
\begin{equation}
G_1^{\mathrm{CI}^{\prime}}(\gamma)
=
\langle
\mathrm{CI}^{\prime}
|
e^{-i\gamma A_1}He^{i\gamma A_1}
|
\mathrm{CI}^{\prime}
\rangle.
\label{eq.VIC.01}
\end{equation}
The second derivative of $G_1^{\mathrm{CI}^{\prime}}$ with respect to $\gamma$ is 
\begin{equation}
\left.\frac{\partial^2 G_1^{\mathrm{CI}^{\prime}}}{\partial \gamma^2}\right|_{\gamma=0}=\sum_{a=1}^8 T_a,
\label{eq.VIC.02}
\end{equation}
where the terms $T_a$ are derived and summarized in Appendix \ref{sec.appendix.derivation of truncated CI gap}. With the choice of overall phase for $|\mathrm{CI}^{\prime}\rangle$ given by Eq. (\ref{eq.appendix.derivation of truncated CI gap.01}), we have 
\begin{eqnarray}
\langle \mathrm{CI}^{\prime}|(A_1)^2|\mathrm{CI}^{\prime}\rangle
&=&
\left(
c_0^2
+
\sum_i c_i^2
+
\sum_{j>i} c_{ij}^2
\right)
\left(
\sum_i y_i^2
\right)
+
4\sum_{j>i} (c_ic_j + c_0c_{ij} )y_iy_j
\nonumber \\
&&
+
2
\sum_{j>i}c_{ij}
\left(
y_j
\sum_{l\ne i,j}c_{il} y_l
+
y_i
\sum_{l\ne i,j}c_{jl} y_l
\right),
\label{eq.VIC.03}
\end{eqnarray}
where in the sum $\sum_{l\ne i,j}$ if $l<i$ then $c_{il}=c_{li}$ (similarly for $c_{jl}$). Eq. (\ref{eq.VIC.03}) gives the constraint $\langle \mathrm{CI}^{\prime}|(A_1)^2|\mathrm{CI}^{\prime}\rangle=1$ when minimizing Eq. (\ref{eq.VIC.02}) with respect to $y_i$. Finally, the truncated CI gap is 
\begin{equation}
\Delta_1^{\mathrm{CI}^{\prime}}
=
\frac{1}{2}
\frac{1}{\langle \mathrm{CI}^{\prime}|(A_1)^2|\mathrm{CI}^{\prime}\rangle}
\left.\frac{\partial^2 G_1^{\mathrm{CI}^{\prime}}}{\partial \gamma^2}\right|_{\gamma=0}.
\label{eq.VIC.04}
\end{equation}

\begin{figure}[h]
\begin{center}
\includegraphics[scale=0.75]{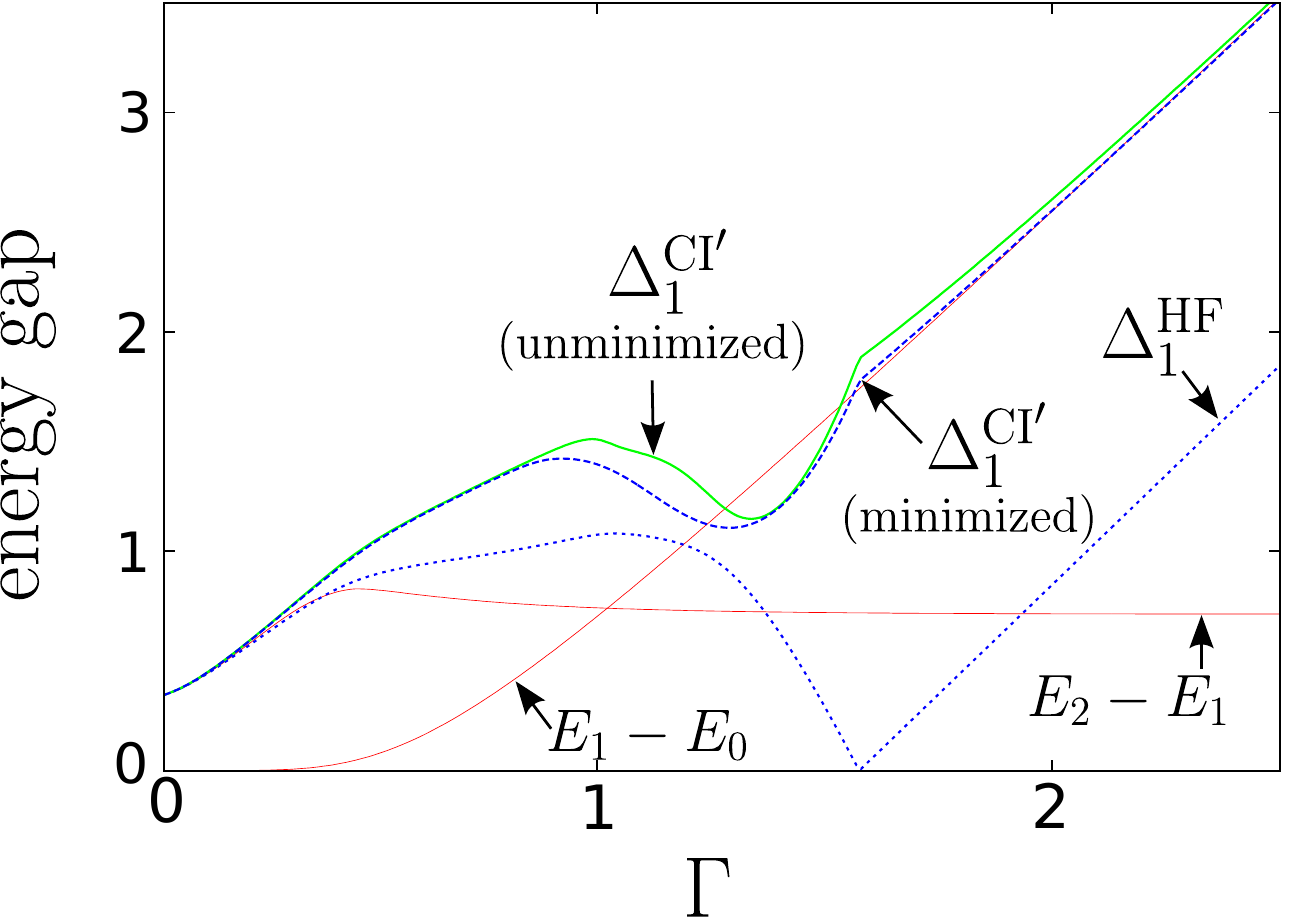}
\caption{(Color online) Energy gaps of the realization of $J_{ij}$ with $\nu=1$ shown in Fig. \ref{fig.Fig10}(a). The curve for $\Delta_1^{\mathrm{CI}^{\prime}}$ (unminimized) (solid, green line) is obtained by substituting the $y_i$ from HF approximation directly into Eq. (\ref{eq.VIC.04}). The curve for $\Delta_1^{\mathrm{CI}^{\prime}}$ (minimized) (dashed, blue line) is obtained by minimizing Eq. (\ref{eq.VIC.02}) subjected to the constraint $\langle\mathrm{CI}^{\prime}|(A_1)^2|\mathrm{CI}^{\prime}\rangle=1$.}
\label{fig.Fig13}
\end{center}
\end{figure}

We computed $\Delta_1^{\mathrm{CI}^{\prime}}$ in two ways. In the first way the $y_i$ from the minimization of $\Delta_1^{\mathrm{HF}}$ is substituted directly into Eq. (\ref{eq.VIC.04}). As this $y_i$ is not the optimal solution, Eq. (\ref{eq.VIC.02}) is not mimimized and Eq. (\ref{eq.VIC.03}) deviates slightly from unity. In the second way, we minimize Eq. (\ref{eq.VIC.02}) subjected to the constraint $\langle \mathrm{CI}^{\prime}|(A_1)^2|\mathrm{CI}^{\prime}\rangle=1$ \cite{note08}. Fig. \ref{fig.Fig13} shows the results for the realization of $J_{ij}$ of Fig. (\ref{fig.Fig10})(a) ($\nu=1$). The $\Delta_1^{\mathrm{CI}^{\prime}}$ calculated in the first and second way are labelled ``unminimized" and ``minimized", respectively. In the ordered regime ($\Gamma<0.5$), the results of $\Delta_1^{\mathrm{HF}}$ and the two $\Delta_1^{\mathrm{CI}^{\prime}}$ are almost identical because the gap is already well reproduced by $\Delta_1^{\mathrm{HF}}$. In the paramagnetic regime ($\Gamma>1.5$), there is significant improvement and the two $\Delta_1^{\mathrm{CI}^{\prime}}$ curves are nearly coincident with the actual gap $E_1-E_0$. In fact, the $y_i$ from HF approximation is already quite close to the optimum, and there is only a slight difference between the two $\Delta_1^{\mathrm{CI}^{\prime}}$ curves in the region $1<\Gamma<2$. Hence, one can compute just the unminimized version of $\Delta_1^{\mathrm{CI}^{\prime}}$ and still obtain accurate values of the gap in both the ordered and paramagnetic regimes. 

In the intermediate regime $0.5<\Gamma<1.5$ where the degeneracy of the ground-state is gradually lifted, it is difficult to pinpoint exactly where the gap goes from being $E_2-E_1$ to being $E_1-E_0$. In this regime, the method proposed here may not be applicable and perhaps a full quantum treatment is necessary.

\begin{figure}[h]
\begin{center}
\includegraphics[scale=0.75]{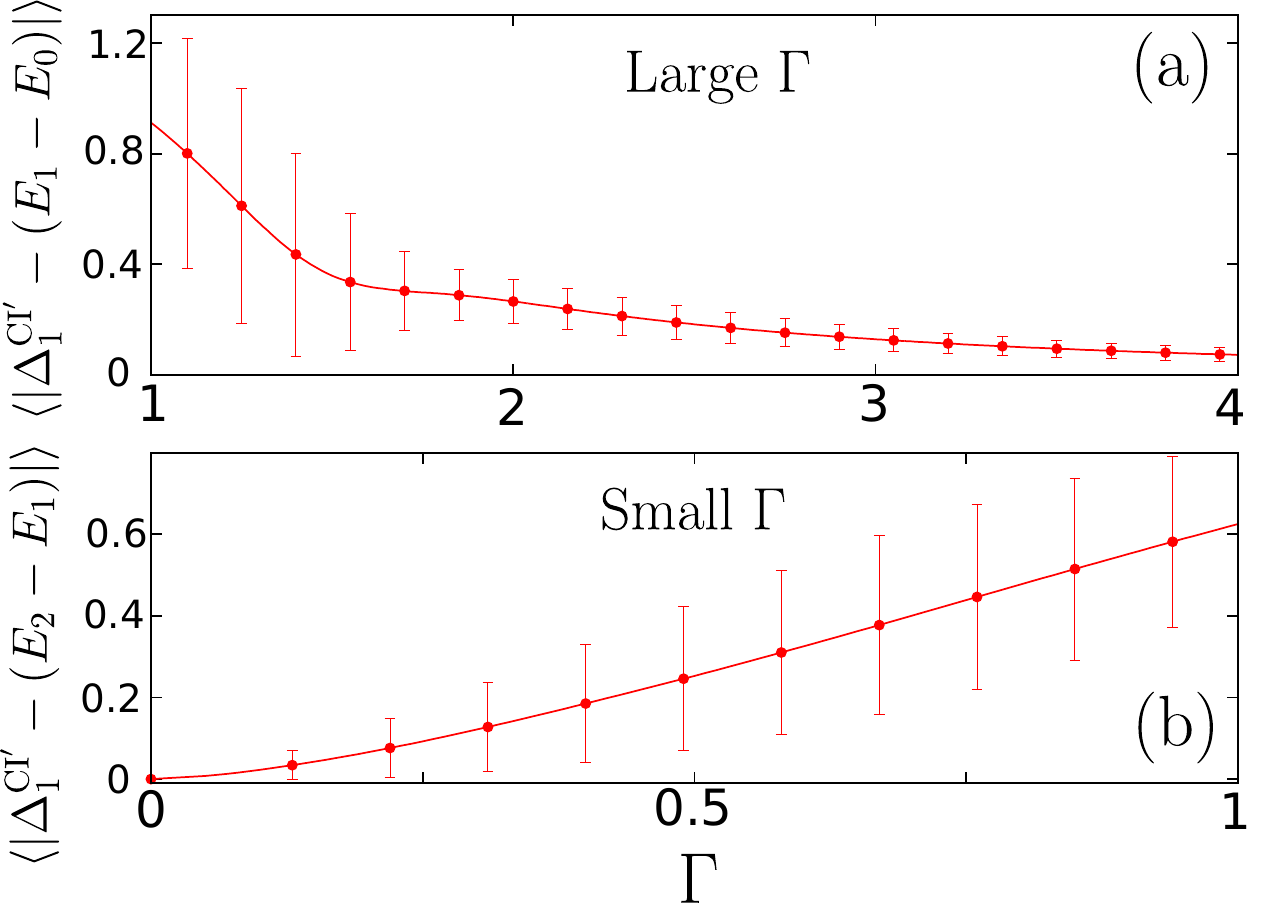}
\caption{(Color online) Absolute error of the energy gap $\Delta_1^{\mathrm{CI}^{\prime}}$(unminimized) averaged over different realizations of $J_{ij}$ with $\nu=1$ for $N=14$. (a) For large $\Gamma$ where the average absolute error is defined as $\langle |\Delta_1^{\mathrm{CI}^{\prime}} -(E_1-E_0)|  \rangle$. (b) For small $\Gamma$ where it is defined as $\langle |\Delta_1^{\mathrm{CI}^{\prime}} -(E_2-E_1)|  \rangle$.  Error bars indicate standard deviation.}
\label{fig.Fig14}
\end{center}
\end{figure}

Fig. \ref{fig.Fig14} shows the error of $\Delta_1^{\mathrm{CI}^{\prime}}$ averaged over different realizations of $J_{ij}$ with $\nu=1$ for $N=14$ \cite{note06}. The unminimized version of $\Delta_1^{\mathrm{CI}^{\prime}}$ is used. For large $\Gamma$ ($>1$), we computed the average absolute error $\langle |\Delta_1^{\mathrm{CI}^{\prime}}-(E_1-E_0)|\rangle$ where comparison is made with $E_1-E_0$. For small $\Gamma$ ($<1$), we computed $\langle |\Delta_1^{\mathrm{CI}^{\prime}}-(E_2-E_1)|\rangle$ where comparison is made with $E_2-E_1$. One sees that as $\Gamma\rightarrow\infty$ and $\rightarrow 0$, the errors (and their fluctuations) decrease to zero.

To conclude this section, we note that for the ferromagnetic model the energy gap for HF approximation and truncated CI can be computed analytically in the thermodynamic limit. These results are summarized in Appendix \ref{sec.appendix.ferromagnetic model.gap}.

\section{Summary and discussions}
\label{sec.discussion}

This paper presents an extensive study of the Sherrington-Kirkpatrick model in transverse field. We propose using the theoretical framework of Hartree-Fock approximation and Configuration Interaction for the simulation of quantum spin-glasses. The main idea is that low-lying spin excitations can account for much of the energies of the ground and first excited states of the  system. A truncated CI basis consisting of one-spin and two-spin flipped states is therefore proposed, thereby avoiding the use of the full Hilbert space. Detailed numerical studies of the ground-state energy and energy gap of the SK model are performed. A novel formula for computing the energy gap is also proposed. The scaling with system size of (i) the energy gap at the critical point and (ii) the leading correction to the extensive part of the ground-state energy are two new insights obtained with our approach.

The dimension of the truncated CI basis scales as $O(N^2)$, much smaller than $2^N$ which is required for full diagonalization. Nevertheless, this gain does not necessarily imply a drastic loss of accuracy. For instance, Fig. \ref{fig.Fig13} shows that the error incurred in the energy gap in the paramagnetic phase is actually very small. Furthermore, this method is simple to implement. There is just a small number of matrix elements and force terms (c.f. Appendices \ref{sec.appendix.matrix elements}  and \ref{sec.appendix.derivation of truncated CI gap}) that needs to be derived analytically and hard coded into the program. These derived terms are then directly applicable for any parameter values of $\Gamma$ and $J_{ij}$. The usage is simpler compared to that of QMC, where usually one needs to first run a few trials to locate the critical point and also to estimate the number of Monte Carlo steps required to converge the data at different parameter values. Moreover, near the critical point it can be computationally expensive to achieve numerical convergence for large system sizes using QMC\cite{Young08,Young10}. These issues do not arise in our proposed method.

In a spin-glass, especially in the thermodynamic limit, there exists many local minima whose energies are very close to each other, but are separated in configuration space by a macroscopic number of spin flips. The one-spin and two-spin flipped states provide a basis to expand the wavefunction centered around the global minimum. By truncating the CI expansion after the two-spin flipped states, we are effectively ignoring the other local minima (and the states around them) even though their energies can be very close to that of the global minimum. This truncation is a valid approximation when the transverse field is not too strong. When $\Gamma$ is turned on from zero, the ground-state wavefunction---a delta function located at the global minimum---starts to acquire a finite width. As the relaxation of this wavepacket is localized around the vicinity of the global minimum, there is almost no overlap with the wavefunctions at other local minima many spin-flips away. Hence, truncated CI is sufficient to describe the ground-state during this initial stage. As $\Gamma$ increases further, however, the effects of tunneling becomes important. Quantum fluctuations now enable the wavepacket at the global minimum to tunnel across energy barriers and superpose with the wavepackets at other local minima, giving rise to a ground-state that is delocalized in configuration space. In particular, as $\Gamma$ approaches criticality, it becomes necessary to include states with multiple spin-flips, and truncated CI is no longer accurate. Indeed, Fig. 
 \ref{fig.Fig07} shows that the error incurred in the ground-state energy goes to zero in the classical limit ($\Gamma\rightarrow0$) and peaks around $\Gamma\approx1.5$ where the system undergoes a change of phase. Hence, an interesting challenge in the numerical simulation of quantum spin-glasses would be the development of a technique that caters just specifically to the regime near criticality. One can then arrive at the full picture by patching together results of different regimes (i.e., paramagnetic, spin-glass, and critical), each obtained using an appropriate method.

In our study of the energy gap, the one-spin flip operator $A_1$ is used for all $\Gamma$. When $\Gamma$ is large (paramagnetic regime), this seems to be a reasonable assumption as we have seen that the gap can be accurately calculated by Eq. (\ref{eq.VIC.04}). When $\Gamma$ is small (ordered regime), however, not all realizations of $J_{ij}$ are $\nu=1$. Some interesting questions arise. Firstly, for a particular realization of $J_{ij}$ drawn from, say, the gaussian distribution Eq. (\ref{eq.II.02}), is there any way to determine its $\nu$ without comparing the energies of all possible spin configurations? But even if the actual $\nu$ is unknown, one can still compare the gap computed using $A_1$ and some other operators and choose the smaller of the two gaps. For $\nu=2$, one can consider
\begin{equation}
A_2
=
\sum_{j>i} y_{ij}\sigma_i^y \sigma_j^y.
\label{eq.summary_discussion.01}
\end{equation}  
It would be interesting to apply the method presented in this paper to these more complex scenarios in future work. 

The results of the HF energy gap shown in Fig. \ref{fig.Fig11} raises some interesting questions for the quantum annealing of the SK model. As system size increases the gap decreases to zero in the entire ordered phase, and this might be a problem for quantum annealing within the ordered phase. Hitherto, analyses on the feasibility of quantum annealing have focused on the vanishing of the energy gap at the critical point. Implicit in the Landau-Zener analysis of the avoided crossing is the assumption that the two energy levels become well-separated after the crossing. This might not be a valid assumption for the SK model in the thermodynamic limit. Hence, even though the gap at the critical point of a continuous transition model does not decrease exponentially with system size, the feasibility of quantum annealing even for such models might be affected by the nature of the gap within the ordered phase. Indeed, such a caveat has also been briefly mentioned in a recent Letter by Liu et al. \cite{Liu15}. The authors, however, did not further pursue their line of thought with a concrete example. Fig. \ref{fig.Fig11} can serve as a quantitative illustration of their concern, using the SK model in transverse field as an example.

\begin{acknowledgements}
This work was partly supported by the Biomedical Research Council of A*STAR (Agency for Science, Technology and Research), Singapore.
\end{acknowledgements}

\appendix

\section{Summary of analytic results for ferromagnetic model in the thermodynamic limit}
\label{sec.appendix.ferromagnetic model}

\subsection{Holstein-Primakoff transform}
\label{sec.appendix.ferromagnetic model.HP}

Eq. (\ref{eq.I.01}) is solved by first performing a Holstein-Primakoff transformation to bosonic operators $b$ and $b^{\dagger}$ \cite{Seoane12,Dusuel05,Das06},
\begin{equation}
S^z + i S^y = \sqrt{s-n}\,\,b,\,\,\,\,\, S^z - i S^y =b^{\dagger}\,\, \sqrt{s-n},\,\,\,\,\,S^x =s-n,
\label{eq.HP.01}
\end{equation}
where $S^{\alpha}=\sum_{i}\sigma_i^{\alpha}$, $n=b^{\dagger}b$, and $s$ is the angular momentum quantum number. One then diagonalize the transformed $H_{\mathrm{ferro}}$ by first expanding in powers of $N$, followed by a Bogoliubov transformation to new operators $\gamma$ and $\gamma^{\dagger}$ to obtain,
\begin{equation}
H_{\mathrm{ferro}}=
\left\{
\begin{array}{lcc}
 -N \Gamma^{\prime} + \sqrt{\Gamma^{\prime}(\Gamma^{\prime}-2J^{\prime})} - \Gamma^{\prime} +2\sqrt{\Gamma^{\prime}(\Gamma^{\prime}-2J^{\prime})}\,\, \gamma^{\dagger}\gamma + O(N^{-1}) & \mathrm{if} & \Gamma^{\prime} \ge2J^{\prime}, \\
 -N \frac{(2J^{\prime})^2+(\Gamma^{\prime})^2}{4J^{\prime}} + \sqrt{(2J^{\prime})^2-(\Gamma^{\prime})^2} - 2J^{\prime}  +2\sqrt{(2J^{\prime})^2-(\Gamma^{\prime})^2}\,\, \gamma^{\dagger}\gamma + O(N^{-1}) & \mathrm{if} & \Gamma^{\prime} <2J^{\prime}, \\
\end{array}
\right.
\label{eq.HP.02}
\end{equation}
where $\Gamma^{\prime}\ge 2J^{\prime}$ ($\Gamma^{\prime}< 2J^{\prime}$) is the paramagnetic (ferromagnetic) phase. In Eq. (\ref{eq.HP.02}), the first term proportional to $N$ is the extensive part of the ground-state energy. It is also obtainable by mean field theory \cite{Das06}. The second term is the leading correction to the extensive part. The coefficient of $\gamma^{\dagger}\gamma$ gives the energy gap. The two latter terms are of order $O(1)$ and are very small compared to the extensive term.

\subsection{Ground-state energy: HF approximation and truncated CI}
\label{sec.appendix.ferromagnetic model.ground-state}

For the ferromagnetic model, the HF equation is
\begin{equation}
\left(2\alpha^2-1\right)
\left[
1
-
\frac{4J^{\prime}}{\Gamma^{\prime}}
\cdot
\frac{N-1}{N}
\cdot
\alpha\sqrt{1-\alpha^2}
\right]
=
0.
\label{eq.appendix.ferro groundstate energy.01}
\end{equation}
We consider the limit $N\rightarrow\infty$. For $\Gamma^{\prime}\ge2J^{\prime}$, the paramagnetic solution is $\alpha=\frac{1}{\sqrt{2}}$. For $\Gamma^{\prime}<2J^{\prime}$, there are two ferromagnetic solutions,
\begin{equation}
\alpha_{\pm}
=
\sqrt{
\frac{1\pm\sqrt{1-\left(\frac{\Gamma^{\prime}}{2J^{\prime}}\right)^2  }}
{2}
}.
\label{eq.appendix.ferro groundstate energy.02}
\end{equation}
The solutions ${\alpha_+\choose \sqrt{1-(\alpha_+)^2}}$ and ${\alpha_-\choose \sqrt{1- (\alpha_-)^2}}$ are related to each other by a spin flip. The HF ground-state energy is
\begin{equation}
E_{\mathrm{ferro}}^{\mathrm{HF}}
=
\left\{
\begin{array}{lcc}
 -N \Gamma^{\prime} -J^{\prime} +O(N^{-1}) & \mathrm{if} & \Gamma^{\prime} \ge2J^{\prime}, \\
 -N \frac{(2J^{\prime})^2+(\Gamma^{\prime})^2}{4J^{\prime}} -\frac{(\Gamma^{\prime})^2}{4J^{\prime}}  +O(N^{-1}) & \mathrm{if} & \Gamma^{\prime} <2J^{\prime}. \\
\end{array}
\right.
\label{eq.appendix.ferro groundstate energy.03}
\end{equation}
Comparing with Eq. (\ref{eq.HP.02}), we see that the first term of HF approximation recovers the extensive part of the ground-state energy.  

\begin{figure}[h]
\begin{center}
\includegraphics[scale=0.75]{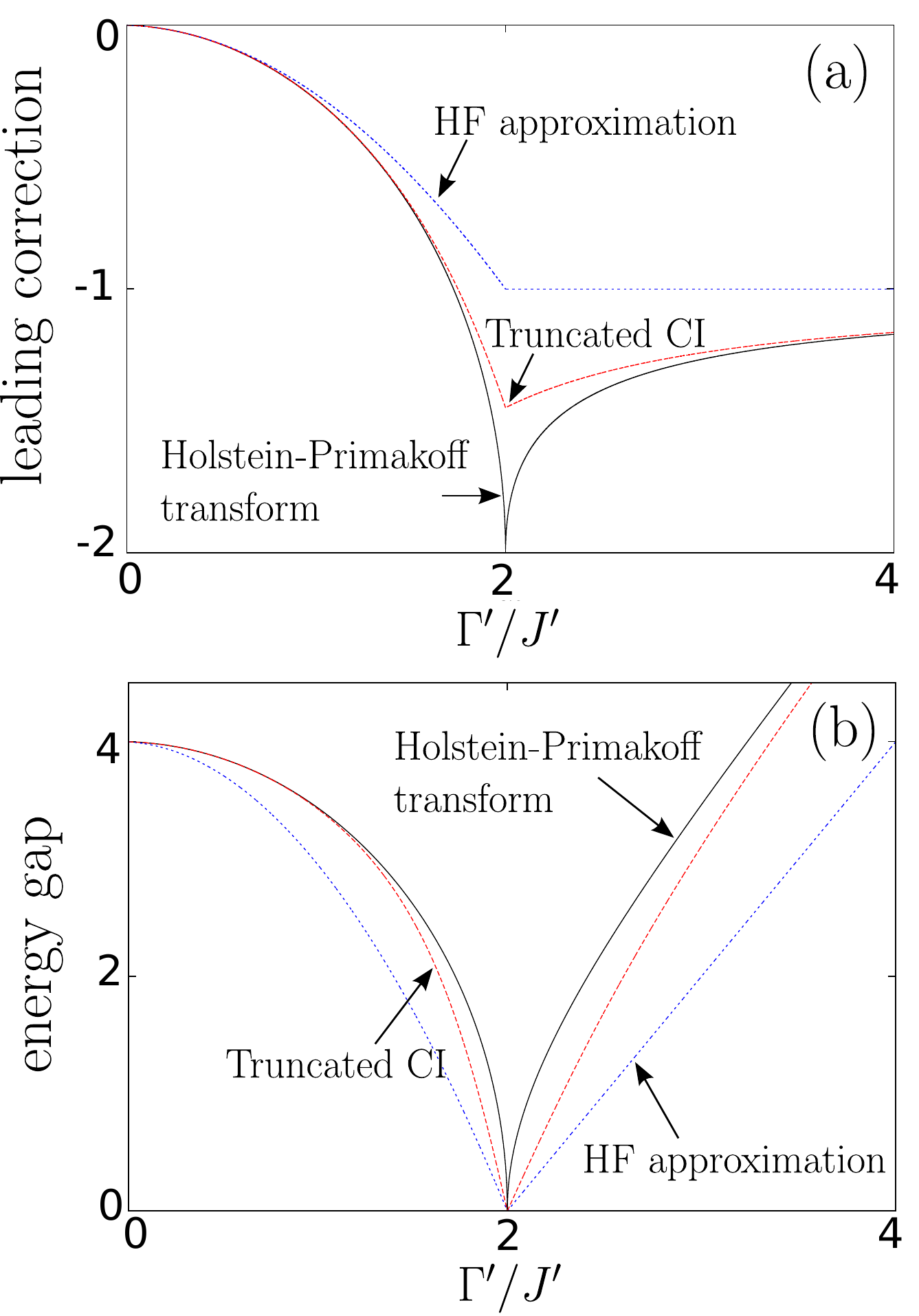}
\caption{(Color online) Comparison of analytic results for the ferromagnetic model Eq. (\ref{eq.I.01}) in the thermodynamic limit, between Holstein-Primakoff transform (black, solid line), HF approximation (blue, dotted line), and truncated CI (red, dashed line). (a) Leading correction to the extensive part of the ground-state energy. (b) Energy gap between the ground and first excited state.}
\label{fig.FigA1}
\end{center}
\end{figure}

To incorporate the effects of one-spin and two-spin flips, we compute the lowest eigenvalue of the truncated CI matrix. In the limit $N\rightarrow\infty$, this can be done analytically. The result is,
\begin{equation}
E_{\mathrm{ferro}}^{\mathrm{CI^\prime}}
=
\left\{
\begin{array}{lcc}
 -N \Gamma^{\prime} -J^{\prime} +2(\Gamma^{\prime}-J^{\prime})-\sqrt{4(\Gamma^{\prime}-J^{\prime})^2+2(J^{\prime})^2}+O(N^{-1}) & \mathrm{if} & \Gamma^{\prime} \ge2J^{\prime}, \\
 -N \frac{(2J^{\prime})^2+(\Gamma^{\prime})^2}{4J^{\prime}} -\frac{(\Gamma^{\prime})^2}{4J^{\prime}}  + 4J^{\prime}-\frac{(\Gamma^{\prime})^2}{2J^{\prime}} -\sqrt{  \left(4J^{\prime}-\frac{(\Gamma^{\prime})^2}{2J^{\prime}}\right)^2 + \frac{(\Gamma^{\prime})^4}{8(J^{\prime})^2} }  +O(N^{-1}) & \mathrm{if} & \Gamma^{\prime} <2J^{\prime}. \\
\end{array}
\right.
\label{eq.appendix.ferro groundstate energy.04}
\end{equation}
A summary of the leading correction to the extensive part of the ground-state energy given by Holstein-Primakoff transform, HF approximation, and truncated CI is shown in Fig. \ref{fig.FigA1}(a). The energies given by HF approximation and truncated CI are both higher than the true ground-state energy, as required by the variational principle.

\subsection{Energy gap: HF approximation and truncated CI}
\label{sec.appendix.ferromagnetic model.gap}

In the HF approximation, the energy gap for the ferromagnetic model is
\begin{equation}
\Delta_{1,\mathrm{ferro}}^{\text{HF}}
=
\left\{
\begin{array}{ccc}
2(\Gamma^{\prime}-2J^{\prime})  +O(N^{-1}) & \text{if} & \Gamma^{\prime}\ge2J^{\prime}, \\
4J^{\prime}-\frac{(\Gamma^{\prime})^2}{J^{\prime}}  +O(N^{-1}) & \text{if} & \Gamma^{\prime}<2J^{\prime}. \\
\end{array}
\right.
\label{eq.appendix.ferro gap.01}
\end{equation}
For truncated CI, the energy gap in the paramagnetic phase ($\Gamma^{\prime}\ge2J^{\prime}$) is
\begin{equation}
\Delta_{1,\mathrm{ferro}}^{\mathrm{CI}^{\prime}}
=
2(\Gamma^{\prime}-2J^{\prime})
\cdot
\frac
{(J^{\prime})^2 + 2(\Gamma^{\prime}-J^{\prime})^2 -(\Gamma^{\prime}-J^{\prime})\sqrt{4(\Gamma^{\prime}-J^{\prime})^2+2(J^{\prime})^2}}
{(J^{\prime})^2+2\Gamma^{\prime} J^{\prime}+10(\Gamma^{\prime}-J^{\prime})^2-(5\Gamma^{\prime}-4J^{\prime})\sqrt{4(\Gamma^{\prime}-J^{\prime})^2+2(J^{\prime})^2}}
+O(N^{-1}).
\label{eq.appendix.ferro gap.02}
\end{equation}
In the ferromagnetic phase ($\Gamma^{\prime}<2J^{\prime}$), it is
\begin{equation}
\Delta_{1,\mathrm{ferro}}^{\mathrm{CI}^{\prime}}
=
\left(
4J^{\prime}-\frac{(\Gamma^{\prime})^2}{J^{\prime}}
\right)
\cdot
\frac
{(4J^{\prime}-\frac{(\Gamma^{\prime})^2}{2J^{\prime}})^2 + \frac{(\Gamma^{\prime})^4}{8(J^{\prime})^2} - (4J^{\prime}-\frac{(\Gamma^{\prime})^2}{2J^{\prime}})\sqrt{(4J^{\prime}-\frac{(\Gamma^{\prime})^2}{2J^{\prime}})^2+\frac{(\Gamma^{\prime})^4}{8(J^{\prime})^2}} }
{5(4J^{\prime}-\frac{(\Gamma^{\prime})^2}{2J^{\prime}})^2+\frac{(\Gamma^{\prime})^4}{8(J^{\prime})^2} + 2(\Gamma^{\prime})^2-2(10J^{\prime}-\frac{(\Gamma^{\prime})^2}{J^{\prime}})\sqrt{(4J^{\prime}-\frac{(\Gamma^{\prime})^2}{2J^{\prime}})^2 + \frac{(\Gamma^{\prime})^4}{8(J^{\prime})^2}}}
+O(N^{-1}).
\label{eq.appendix.ferro gap.03}
\end{equation}
A summary of the energy gap given by Holstein-Primakoff transform, HF approximation, and truncated CI is shown in Fig. \ref{fig.FigA1}(b).

\section{Matrix elements of truncated CI matrix}
\label{sec.appendix.matrix elements}

\begin{eqnarray}
\langle 0|H|0 \rangle
&=&
E^{\mathrm{HF}}
\\
\langle 0|H|i \rangle
&=&
2i\alpha_i\beta_i\sum_{a\ne i}J_{ia}(\alpha^2_a-\beta_a^2) - i\Gamma(\alpha_i^2-\beta_i^2).
\\
\langle 0|H|ij \rangle
&=&
4J_{ij} \alpha_i\beta_i\alpha_j\beta_j.
\\
\langle i|H|j \rangle
&=&
\left
\{
\begin{array}{ccc}
E^{\mathrm{HF}}
+
4\Gamma \alpha_i\beta_i
&&\\
+
2(\alpha_i^2-\beta_i^2)\sum_{a\ne i} J_{ia}(\alpha_a^2-\beta_a^2)
&\text{if}& \langle i|H|i \rangle,  \\
-4J_{ij} \alpha_i\beta_i\alpha_j\beta_j    &\text{otherwise}.&   \\
\end{array}
\right.
\\
\langle k|H|ij \rangle
&=&
\left
\{
\begin{array}{ccc}
2i\alpha_j\beta_j\sum_{a\ne i,j} J_{ja}(\alpha_a^2-\beta_a^2) && \\
-2iJ_{ij}\alpha_j\beta_j(\alpha_i^2-\beta_i^2)-i\Gamma(\alpha_j^2-\beta_j^2)    &\text{if}& \langle i|H|ij \rangle,  \\
  0  &\text{otherwise}.&   \\
\end{array}
\right.
\\
\langle kl|H|ij \rangle
&=&
\left
\{
\begin{array}{ccc}
E^{\mathrm{HF}}+4\Gamma(\alpha_i\beta_i+\alpha_j\beta_j)
   && \\
+2(\alpha_i^2-\beta_i^2)\sum_{a\ne i,j}J_{ia}(\alpha_a^2-\beta_a^2)       && \\
+2(\alpha_j^2-\beta_j^2)\sum_{a\ne j,i}J_{ja}(\alpha_a^2-\beta_a^2) 
    &\text{if}&  \langle ij|H|ij \rangle, \\
-4J_{jl}\alpha_j\beta_j\alpha_l\beta_l    &\text{if}&  \langle il|H|ij \rangle, \\
0    &\text{otherwise}.& \\
\end{array}
\right.
\\ \nonumber
\end{eqnarray}

\section{Derivation of Eq. (\ref{eq.VII.02})}
\label{sec.appendix.derivation HF gap}

Since $e^{i\gamma A_1}$ is factorizable, we have
\begin{eqnarray}
e^{i\gamma A_1}|0\rangle
&=&
\prod_i 
e^{i\gamma y_i\sigma_i^y}
{\alpha_i \choose \beta_i}
\nonumber\\
&=&
\prod_i 
{\alpha_i \cos\gamma y_i + \beta_i \sin\gamma y_i \choose -\alpha_i\sin\gamma y_i + \beta_i \cos\gamma y_i}
\nonumber\\
&=&
\prod_i 
{\bar{\alpha}_i  \choose \bar{\beta}_i},
\nonumber\\
\label{eq.appendix.derivation of hf gap.01}
\end{eqnarray}
where $\bar{\alpha}_i=\alpha_i \cos\gamma y_i + \beta_i \sin\gamma y_i$ and $\bar{\beta}_i=-\alpha_i\sin\gamma y_i + \beta_i \cos\gamma y_i$. Hence,
\begin{equation}
G_1^{\mathrm{HF}}(\gamma)=E^{\mathrm{HF}}(\bar{\bm{\alpha}}(\gamma),\bar{\bm{\beta}}(\gamma)).
\label{eq.appendix.derivation of hf gap.02}
\end{equation}
Differentiating $G_1^{\mathrm{HF}}(\gamma)$ twice with respect to $\gamma$ using chain rule, we have
\begin{equation}
\left.
\frac{\partial^2 G_1^{\mathrm{HF}}}{\partial \gamma^2}
\right|_{\gamma=0}
=
\left[
\sum_{i=1}^{N}
\sum_{j=1}^{N}
y_i
y_j
\hat{\partial}_{ij}
-
\sum_{i=1}^{N}
y_i^2
\hat{\partial}_i
\right]
E^{\mathrm{HF}}(\bar{\bm{\alpha}},\bar{\bm{\beta}}),
\label{eq.appendix.derivation of hf gap.03}
\end{equation}
where
\begin{equation}
\hat{\partial}_i
=
\left.
\left[
\bar{\alpha}_i
\frac{\partial}{\partial \bar{\alpha}_i}
+
\bar{\beta}_i
\frac{\partial}{\partial \bar{\beta}_i}
\right]
\right|_{\gamma=0},
\label{eq.appendix.derivation of hf gap.04}
\end{equation}
and
\begin{equation}
\hat{\partial}_{ij}
=
\left.
\left[
\bar{\alpha}_i\bar{\alpha}_j
\frac{\partial^2}{\partial\bar{\beta}_i\partial\bar{\beta}_j}
-
\bar{\alpha}_j\bar{\beta}_i
\frac{\partial^2}{\partial\bar{\alpha}_i\partial\bar{\beta}_j}
-
\bar{\alpha}_i\bar{\beta}_j
\frac{\partial^2}{\partial\bar{\beta}_i\partial\bar{\alpha}_j}
+
\bar{\beta}_i\bar{\beta}_j
\frac{\partial^2}{\partial\bar{\alpha}_i\partial\bar{\alpha}_j}
\right]
\right|_{\gamma=0}
.
\label{eq.appendix.derivation of hf gap.05}
\end{equation}
Eq. (\ref{eq.appendix.derivation of hf gap.03}) is evaluated by substituting Eq. (\ref{eq.III.04}) into the right side and working out the derivatives. The constraint Eq. (\ref{eq.VI.08}) becomes
\begin{equation}
\langle 0|(A_1)^2|0\rangle=1,
\label{eq.appendix.derivation of hf gap.06}
\end{equation}
and is easily shown to be Eq. (\ref{eq.VII.03}). Inserting Eqs. (\ref{eq.appendix.derivation of hf gap.03}) and (\ref{eq.appendix.derivation of hf gap.06}) into Eq. (\ref{eq.VI.06}), we get Eq. (\ref{eq.VII.02}).

\section{Derivation of the terms in Eq. (\ref{eq.VIC.02})}
\label{sec.appendix.derivation of truncated CI gap}

With an appropriate choice of phase for the first element $c_0$, Eq. (\ref{eq.V.00}) can be written as
\begin{equation}
|\mathrm{CI}^{\prime}\rangle=c_0 |0\rangle + i\sum_{i} c_i |i\rangle + \sum_{j>i} c_{ij} |ij\rangle,
\label{eq.appendix.derivation of truncated CI gap.01}
\end{equation}
where $c_0$, $c_i$, and $c_{ij}$ are all real. In addition, let us denote the matrix elements listed in Appendix \ref{sec.appendix.matrix elements} as   
\begin{equation}
\langle n | H | m\rangle
=
H_{n,m}(\bm{\alpha},\bm{\beta}),
\label{eq.appendix.derivation of truncated CI gap.02}
\end{equation}
where the dependence on $\bm{\alpha}$ and $\bm{\beta}$ are made explicit. For instance, for Eq. (B5), $n=k$ and $m=ij$. In view of Eq. (\ref{eq.appendix.derivation of hf gap.01}), it is apparent that
\begin{equation}
\langle n |e^{-i\gamma A_1} H e^{i\gamma A_1}| m\rangle
=
H_{n,m}(\bar{\bm{\alpha}},\bar{\bm{\beta}}).
\label{eq.appendix.derivation of truncated CI gap.03}
\end{equation}
With the above notations, the right side of Eq. (\ref{eq.VIC.01}) is expanded to give

\begin{eqnarray}
G_1^{\mathrm{CI}^{\prime}}(\gamma)
&=&
\left(c_0^2 + \sum_i c_i^2   + \sum_{j>i} c_{ij}^2\right)
H_{0,0}(\bar{\bm{\alpha}},\bar{\bm{\beta}})
+
2ic_0\sum_i c_i H_{0,i}(\bar{\bm{\alpha}},\bar{\bm{\beta}})
+
2c_0\sum_{j>i}c_{ij}H_{0,ij}(\bar{\bm{\alpha}},\bar{\bm{\beta}})
\nonumber\\
&&
+
\sum_i c_i^2 R_i(\bar{\bm{\alpha}},\bar{\bm{\beta}})
+
2
\sum_{j>i}c_ic_j H_{i,j}(\bar{\bm{\alpha}},\bar{\bm{\beta}})
+
\sum_{j>i}c_{ij}^2 R_{ij}(\bar{\bm{\alpha}},\bar{\bm{\beta}})
-2i\sum_i \sum_{j\ne i}c_ic_{ij} H_{i,ij}(\bar{\bm{\alpha}},\bar{\bm{\beta}})
\nonumber\\
&&
+
\sum_{j>i}c_{ij}\sum_{l\ne i,j}\left[c_{il}H_{ij,il}(\bar{\bm{\alpha}},\bar{\bm{\beta}}) + c_{jl}H_{ij,jl}(\bar{\bm{\alpha}},\bar{\bm{\beta}})  \right],
\label{eq.appendix.derivation of truncated CI gap.04}
\end{eqnarray}
where we let $c_{ij}=c_{ji}$ whenever $i>j$, and $R_n(\bar{\bm{\alpha}},\bar{\bm{\beta}})$ is defined as
\begin{equation}
R_n(\bar{\bm{\alpha}},\bar{\bm{\beta}})
=
H_{n,n}(\bar{\bm{\alpha}},\bar{\bm{\beta}})
-
H_{0,0}(\bar{\bm{\alpha}},\bar{\bm{\beta}}).
\label{eq.appendix.derivation of truncated CI gap.05}
\end{equation}

The second derivative of $G_1^{\mathrm{CI}^{\prime}}$ with respect to $\gamma$ is implemented by the same differential operator in Eq. (\ref{eq.appendix.derivation of hf gap.03}), i.e.,
\begin{equation}
\partial^2_{\gamma}
\equiv
\left.
\frac{\partial^2}{\partial \gamma^2}
\right|_{\gamma=0}
=
\sum_{i=1}^{N}
\sum_{j=1}^{N}
y_i
y_j
\hat{\partial}_{ij}
-
\sum_{i=1}^{N}
y_i^2
\hat{\partial}_i
.
\label{eq.appendix.derivation of truncated CI gap.06}
\end{equation}
Applying $\partial^2_{\gamma}$ to the right side of Eq. (\ref{eq.appendix.derivation of truncated CI gap.04}), noting that it is linear and hence does not affect the coefficients $c_0$, $c_i$, and $c_{ij}$, we see that we need to compute $\partial^2_{\gamma}$ for the matrix elements listed in Appendix \ref{sec.appendix.matrix elements}. The calculation is similar to that of Appendix \ref{sec.appendix.derivation HF gap}. The result is Eq. (\ref{eq.VIC.02}) where
\begin{eqnarray} 
T_1
&=&
\left(c_0^2 + \sum_i c_i^2   + \sum_{j>i} c_{ij}^2\right) \partial^2_{\gamma} H_{0,0}(\bar{\bm{\alpha}},\bar{\bm{\beta}})
\nonumber\\
&=&
\left(c_0^2 + \sum_i c_i^2   + \sum_{j>i} c_{ij}^2\right)
\left(
-16
\sum_{i\ne j} J_{ij} \alpha_i\beta_i \alpha_j\beta_j  y_iy_j
+
2\Gamma \sum_{i} \frac{y_i^2}{\alpha_i\beta_i}
\right)
\\
T_2
&=&
2ic_0\sum_i c_i \partial^2_{\gamma} H_{0,i}(\bar{\bm{\alpha}},\bar{\bm{\beta}})
\nonumber\\
&=&
32c_0\sum_{i,j}J_{ij} \alpha_i\beta_ic_j(\alpha_j^2 - \beta_j^2) y_iy_j
+
16c_0\sum_{i,j}J_{ij} y_i^2 (\alpha_i^2 - \beta_i^2) c_j \alpha_j \beta_j
\\
T_3
&=&
2c_0\sum_{j>i}c_{ij}\partial^2_{\gamma}H_{0,ij}(\bar{\bm{\alpha}},\bar{\bm{\beta}})
\nonumber\\
&=&
8c_0
\sum_{i,j}
c_{ij}J_{ij}y_iy_j(\alpha_i^2 - \beta_i^2)(\alpha_j^2 - \beta_j^2)
-
32c_0
\sum_{i,j}c_{ij}J_{ij}\alpha_i \beta_i\alpha_j \beta_j y_i^2
\\
T_4
&=&
\sum_i c_i^2 \partial^2_{\gamma} R_i(\bar{\bm{\alpha}},\bar{\bm{\beta}})
\nonumber\\
&=&
64\sum_{i,j}J_{ij}c_i^2y_iy_j\alpha_i \beta_i\alpha_j \beta_j
-
8\sum_{i,j}J_{ij}c_i^2y_j^2(\alpha_i^2 - \beta_i^2)(\alpha_j^2 - \beta_j^2)
-4\Gamma\sum_i \frac{c_i^2 y_i^2}{\alpha_i \beta_i}
\\
T_5
&=&
2
\sum_{j>i}c_ic_j \partial^2_{\gamma} H_{i,j}(\bar{\bm{\alpha}},\bar{\bm{\beta}})
\nonumber \\
&=&
-8\sum_{i\ne j}J_{ij}c_ic_jy_iy_j(\alpha_i^2 - \beta_i^2)(\alpha_j^2 - \beta_j^2)
+
32\sum_{i\ne j}J_{ij}c_ic_j\alpha_i \beta_i\alpha_j \beta_j y_i^2
\\
T_6
&=&
\sum_{j>i}c_{ij}^2  \partial^2_{\gamma} R_{ij}(\bar{\bm{\alpha}},\bar{\bm{\beta}})
\nonumber\\
&=&
-
4\Gamma\sum_{i\ne j}\frac{c_{ij}^2y_i^2}{\alpha_i \beta_i}
-
64\sum_{i\ne j}J_{ij}c_{ij}^2y_iy_j\alpha_i \beta_i\alpha_j \beta_j
+
16\sum_{i\ne j}J_{ij}c_{ij}^2(\alpha_i^2 - \beta_i^2)(\alpha_j^2 - \beta_j^2)y_i^2
\nonumber\\
&&
+
64\sum_{i\ne j}c_{ij}^2y_i\alpha_i \beta_i\sum_{u}J_{iu}y_u\alpha_u \beta_u
-
8\sum_{i\ne j}c_{ij}^2(\alpha_i^2 - \beta_i^2)\sum_{u}J_{iu}y_u^2(\alpha_u^2 - \beta_u^2)
\\
T_7
&=&
-2i\sum_i \sum_{j\ne i}c_ic_{ij} \partial^2_{\gamma} H_{i,ij}(\bar{\bm{\alpha}},\bar{\bm{\beta}})
\nonumber\\
&=&
-
32\sum_{i\ne j}c_ic_{ij}y_j(\alpha_j^2 - \beta_j^2)\sum_{u}J_{ju}y_u\alpha_u \beta_u
+
64\sum_{i\ne j}c_ic_{ij}J_{ij} y_iy_j\alpha_i \beta_i(\alpha_j^2 - \beta_j^2)
\nonumber\\
&&
+
32\sum_{i\ne j}c_ic_{ij}J_{ij}\alpha_j \beta_j (\alpha_i^2 - \beta_i^2) (y_i^2+y_j^2)
-
16\sum_{i\ne j}c_ic_{ij}\alpha_j \beta_j \sum_{u}J_{ju}y_u^2(\alpha_u^2 - \beta_u^2)
\\
T_8
&=&
\sum_{j>i}c_{ij}\sum_{l\ne i,j}\left[c_{il} \partial^2_{\gamma} H_{ij,il}(\bar{\bm{\alpha}},\bar{\bm{\beta}}) + c_{jl} \partial^2_{\gamma} H_{ij,jl}(\bar{\bm{\alpha}},\bar{\bm{\beta}})  \right]
\nonumber \\
&=&
-8\sum_{j>i}
c_{ij}
\left\{
\sum_{l\ne i,j} c_{il} J_{lj} \left[ y_ly_j (\alpha_l^2 - \beta_l^2)(\alpha_j^2 - \beta_j^2) -2 \alpha_l \beta_l \alpha_j \beta_j (y_l^2+y_j^2) \right]
\right.
\nonumber\\
&&
\left.
+
\sum_{l\ne i,j} c_{jl} J_{li}
\left[ y_ly_i (\alpha_l^2 - \beta_l^2)(\alpha_i^2 - \beta_i^2) -2 \alpha_l \beta_l \alpha_i \beta_i (y_l^2+y_i^2) \right]
\right\}
\end{eqnarray}

\begin{thebibliography}{99}
\bibitem{note.intro.01} See, for example, Refs. \cite{Seoane12}, \cite{Dusuel05}, and \cite{Das06} for the Holstein-Primakoff transform. The results of the Holstein-Primakoff transform for the ferromagnetic model Eq. (\ref{eq.I.01}) is presented in Appendix \ref{sec.appendix.ferromagnetic model.HP}.
\bibitem{note.intro.02} Details on CI can be found in standard texts on quantum chemistry such as Ref. \cite{Jensen07}.
\bibitem{note02} For $N=8$ to 14. Each $N$ is averaged over 10000 realizations of $J_{ij}.$
\bibitem{note05} We solve the mean-field equations (replica-symmetric ansatz) of Ref. \cite{Thirumalai89} in the limit $\beta\rightarrow\infty$ (i.e. $T=0$).
\bibitem{note05.5} Note that in the ordered regime, the first excited state energy is $E_2$ because the ground-state energy is doubly-degnerate (i.e. $E_0=E_1$).
\bibitem{note07} This can be evinced from the inset of Fig. \ref{fig.Fig10}(a).
\bibitem{note08} We consider the Lagrangian $\mathcal{L}(\vec{y})=\frac{\partial^2 G_1^{\mathrm{CI}^{\prime}}}{\partial \gamma^2} - \lambda(\langle \mathrm{CI}^{\prime}|(A_1)^2|\mathrm{CI}^{\prime}\rangle -1)$ where $\lambda$ is a Lagrange multiplier. As the optimal solution lies on a saddle point, we minimize instead the norm of the gradient of $\mathcal{L}$, which requires the derivatives of Eqs. (\ref{eq.VIC.02}) and  (\ref{eq.VIC.03}) with respect to $y_i$.
\end{thebibliography}
\end{document}